\documentclass[twocolumn, tighten]{openjournal}
\usepackage{url}
\usepackage{xcolor}
\definecolor{xlinkcolor}{cmyk}{1,1,0,0}
\usepackage[
 bookmarks=true, 
 pdfnewwindow=true,      
 colorlinks=true,    
 linkcolor=xlinkcolor,     
 citecolor=xlinkcolor,     
 filecolor=xlinkcolor,  
 urlcolor=xlinkcolor,      
 final=true,
]{hyperref}

\usepackage{orcidlink}
\usepackage[english]{babel}
\usepackage[babel=true]{csquotes}
\usepackage{booktabs}
\usepackage{threeparttable}
\usepackage{tabularx}

\usepackage{multirow}
\newcolumntype{Y}{>{\centering\arraybackslash}m{1cm}}

\makeatletter
\renewcommand\@makecaption[2]{%
  \par
  \vskip\abovecaptionskip
  \begingroup
    \footnotesize\rmfamily
    \begingroup
      \samepage
      \flushing
      \let\footnote\@footnotemark@gobble
      \ifnum\pdfstrcmp{\@captype}{table}=0
        \@make@capt@title{\textsc{Table \thetable}}{#2}%
      \else
        \ifnum\pdfstrcmp{\@captype}{figure}=0
          \@make@capt@title{\textsc{Figure \thefigure}}{#2}%
        \else
          \@make@capt@title{#1}{#2}%
        \fi
      \fi\par
    \endgroup
  \endgroup
  \vskip\belowcaptionskip
}
\makeatother

\usepackage{etoolbox}
\AtBeginEnvironment{tabular}{\renewcommand{\arraystretch}{1.5}}
\AtBeginEnvironment{tabularx}{\renewcommand{\arraystretch}{1.5}}
\usepackage{graphicx}
\graphicspath{{./}{figures/}}
\usepackage{bm}
\usepackage{xspace}

\usepackage[shortlabels]{enumitem}

\newcommand{\cosmicweb}{\textsc{cosmICweb}\xspace}
\newcommand{\cosmicweburl}{https://cosmicweb.eu}
\newcommand{\cosmicwebmusic}{\texttt{cosmicweb-music}\xspace}

\newcommand{\music}{\textsc{music}\xspace}
\newcommand{\panphasia}{\text{Panphasia}\xspace}
\newcommand{\icgen}{\textsc{ic\_gen}\xspace}
\newcommand{\rockstar}{\textsc{rockstar}\xspace}
\newcommand{\consistenttrees}{\textsc{consistent-trees}\xspace}

\newcommand{\get}{\texttt{GET}\xspace}
\newcommand{\post}{\texttt{POST}\xspace}

\newcommand{\agora}{\textsc{agora}\xspace}
\newcommand{\rhapsody}{\textsc{Rhapsody}\xspace}
\newcommand{\rhapsodypc}{\textsc{Rhapsody\_PlanckCosmo}\xspace}
\newcommand{\eagle}{\textsc{Eagle}\xspace}
\newcommand{\flamingo}{\textsc{Flamingo}\xspace}

\newcommand{\hMsun}{\(h^{-1}\mathrm{M}_\odot\)\xspace}
\newcommand{\hMpc}{\(h^{-1}\mathrm{Mpc}\)\xspace}
\newcommand{\hkpc}{\(h^{-1}\mathrm{kpc}\)\xspace}

\begin{document}

\title{cosmICweb: Cosmological Initial Conditions for Zoom-in Simulations in the Cloud}
\shorttitle{cosmICweb}

\author{
    Michael Buehlmann$^{1,\star}$\orcidlink{0000-0002-8469-4534},
    Lukas Winkler$^2$\orcidlink{0000-0002-6792-6743},
    Oliver Hahn$^{2,3}$\orcidlink{0000-0001-9440-1152},
    John C. Helly$^4$\orcidlink{0000-0002-0647-4755},
    Adrian Jenkins$^4$\orcidlink{0000-0003-4389-2232}
}
\thanks{$^{\star}$E-mail: mbuehlmann@anl.gov}
\shortauthors{Buehlmann, Winkler, Hahn, Helly, \& Jenkins}

\affiliation{$^1$Computational Science Division, Argonne National Laboratory, Lemont, IL 60439, USA}
\affiliation{$^2$Department of Astrophysics, University of Vienna, Türkenschanzstraße 17 (Sternwarte), 1180 Wien, Austria}
\affiliation{$^3$Department of Mathematics, University of Vienna, Oskar-Morgenstern-Platz 1, 1090 Vienna, Austria}
\affiliation{$^4$Institute for Computational Cosmology, Department of Physics, University of Durham, South Road, Durham, DH1 3LE, UK}


\begin{abstract}
We present the online service \cosmicweb (\textbf{COSM}ological \textbf{I}nitial \textbf{C}onditions on the \textbf{WEB})— the first database and web interface to store, analyze, and disseminate initial conditions (ICs) for zoom simulations of objects forming in cosmological simulations: from galaxy clusters to galaxies and more. Specifically, we store compressed information about the Lagrangian proto-halo patches for all objects in a typical simulation merger tree along with properties of the halo/galaxy across cosmic time. This enables a convenient web-based selection of the desired zoom region for an object  fitting user-specified selection criteria. The information about the region can then be used with the \music code to generate the zoom ICs for the simulation. In addition to some other simulations, we currently support all objects in the \eagle simulation database, so that for example the \textsc{Auriga} simulations are easily reproduced, which we demonstrate explicitly. The framework is easily extensible to include other simulations through an API that can be added to an existing database structure and with which \cosmicweb can then be interfaced. We make the web portal and database publicly available to the community. 
\end{abstract}


\section{Introduction}

In cosmology and extragalactic astrophysics, numerical simulations are a fundamental tool to study the growth and evolution of structures in the universe across an enormous range of scales \citep[e.g.,][for recent reviews]{Vogelsberger:2020,AnguloHahn:2022}. On the largest scales up to several gigaparsec, they probe the cosmological large scale structure. On the smallest scales ranging from several parsec to kiloparsec, they can resolve the formation and structure of individual galaxies.
The computational cost of computing and storing the full dynamic range in the entire simulated volume sets a limit on the upper and lower scales that can be simulated in a single run. A common approach to circumvent this limitation is the \emph{multi-mass} or \emph{zoom} technique, in which the immediate region around the object of interest is simulated with a highly increased resolution compared to its surrounding large scale environment \citep{Katz1994, Navarro1994}.
This method allows capturing both the influences of the cosmic environment from fluctuations on scales equal and larger than the object as well as small scale fluctuations affecting the structure of the object directly.
Zoom-in initial conditions can be generated by degrading a high-resolution initial condition outside the zoom region \citep[e.g.,][]{Garaldi2016} or by refining the zoom region \citep[e.g.,][]{Bertschinger2001, Jenkins2010, Hahn2011}. Since the former ``degrading'' approach is computationally limited by memory for deep zoom levels, zoom initial conditions are usually generated using one of the refinement techniques.

When computing such zoom initial conditions, a fundamental ingredient is that the noise field (also known as the \emph{random phases}) is spatially stable without having to generate it at full resolution. This can be achieved in different ways, with the \panphasia approach \citep{Jenkins2013} of an expansion in terms of hierarchically defined basis functions being particularly powerful. \panphasia has been adopted for various large recent simulations such as \eagle \citep{Schaye2015} and \flamingo \citep{Schaye:Flamingo:2023}. Various objects such as e.g., dwarf and Milk-Way-type galaxies in the \textsc{Auriga} simulations \citep{Grand2024} have been re-simulated at higher resolution using zoom-ins on for example the \eagle volume. However, resimulating objects from these simulations so far required proprietary software and access to the particle data at initial and final time. Often, the parent simulations of zoom simulations are not even published in other studies.

\begin{figure*}
    \centering
    \includegraphics[width=0.75\textwidth]{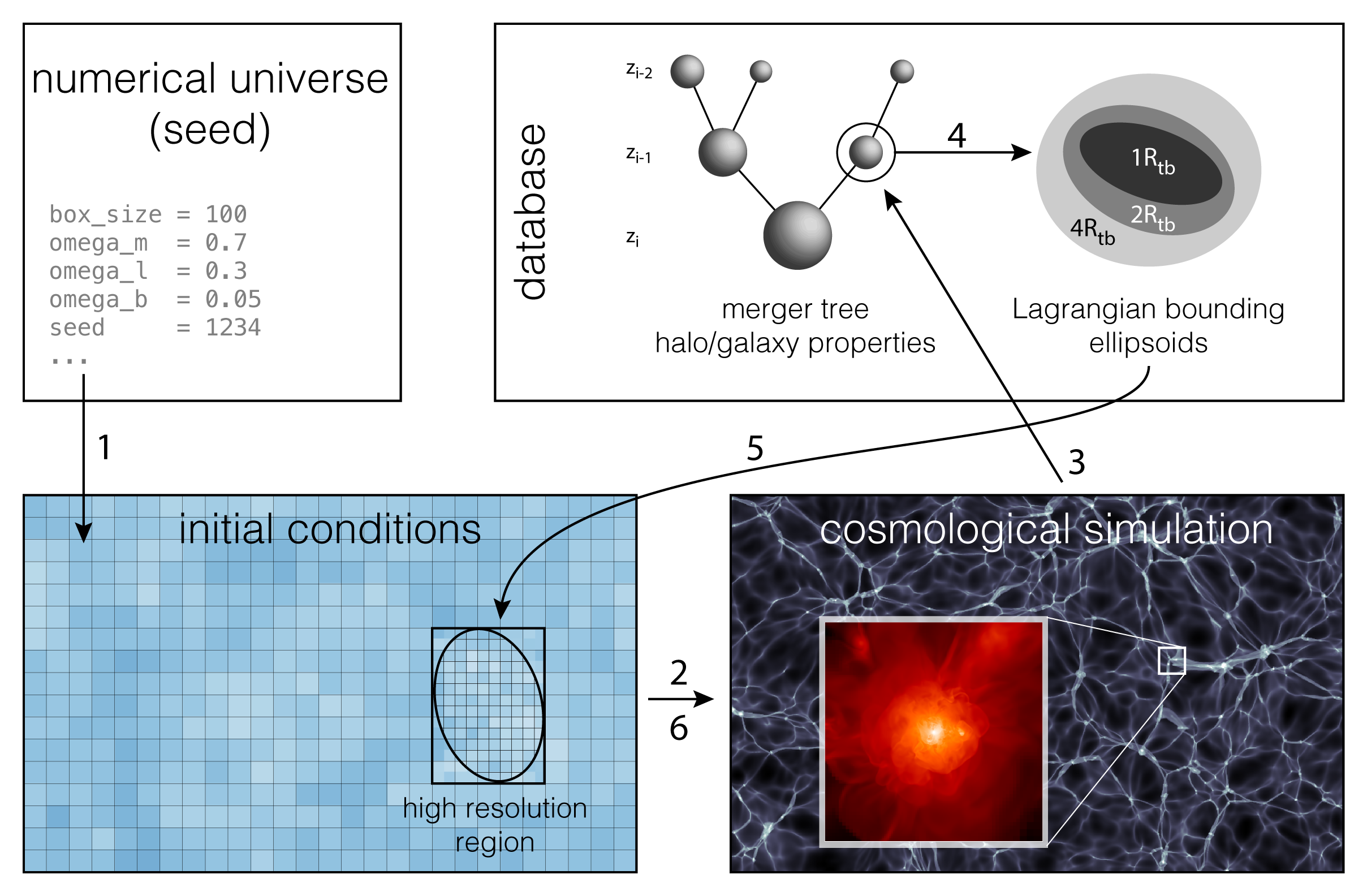}
    \caption{Illustration of the steps involved in the creation of a zoom simulation. A detailed description of the process is provided in the main text. The \cosmicweb database stores halo catalogs, merger trees and the associated Lagrangian proto-halo patches from existing simulations, all accessible over an intuitive web interface.}\label{fig:concept}
\end{figure*}

Once the random field is defined, setting up zoom-in simulations requires knowledge of the initial region which forms the object of interest as the universe evolves, the \emph{proto-halo patch}. 
This Lagrangian volume has to be determined from a (lower resolution) full-box simulation, the ``parent simulation'', in which  one has to find the object of interest and trace its constituent particles back to the initial conditions.
Extra care needs to be taken to not contaminate the zoom region with low-resolution particles, as these can have a significant effect on measured gas densities and substructure formation \citep[see][]{Onorbe2013}.
The typical setup process for zoom-simulations is illustrated in \autoref{fig:concept} and typically consists of the following steps:
\begin{itemize}
    \item[1-3] Setting up, running, and analyzing the parent simulation. The parent simulation is characterized by the white noise field used to generate the initial conditions (i.e. the random number generator and its seed) and the cosmological parameters under which the system is evolved. Various $N$-body codes exist to run the simulation. In order to identify halos or regions of interest in later steps, a halo or a more general structure finder has to be run in-situ or on the simulation output. In order to obtain information on the assembly history of halos, it is desirable to generate frequent outputs and build merger trees that connect halos across time.
    \item[4] Identify target(s): From the merger trees, halo catalogs, or directly from the particle snapshots, select the halos (or a more general region) that match the research goal at hand.
    \item[5] Determine the Lagrangian volume: Back-trace the particles within the target region to their Lagrangian position. In the case of halos, one usually identifies the particles within a ``traceback''-radius who form the proto-halo patch in Lagrangian coordinates.
    \item[6] Create high-resolution initial conditions refined on this patch, using a  multi-scale initial condition generator such as \music \citep{Hahn2011} or \icgen \citep{Jenkins2010, Jenkins2013}.
\end{itemize}

The steps 1 to 5 are preparatory steps that usually have to be repeated for new zoom simulation, unless they are based on the same parent volume. Public halo and galaxy catalogs of large simulations exist \citep[e.g.,][]{McAlpine2016, Klypin:2017, Nelson2019, Heitmann2019}, but they are usually not associated with proto-halo patches. Furthermore, the absence of public access to initial conditions in research articles complicates the direct comparison between different simulation codes, baryonic physics prescriptions, and the verification of scientific results.

With \cosmicweb\footnote{short for ``\textbf{COSM}ological \textbf{I}nitial \textbf{C}onditions on the \textbf{WEB}''}, we address three issues with the current work flow for setting up and running zoom simulations:
\begin{itemize}[itemsep=1pt]
    \item \emph{Finding interesting objects to simulate.} We provide access to databases of pre-run simulations in which the user can search for halos (and galaxies) at various redshifts matching their research project requirements.
    \item \emph{Generating zoom initial conditions.} For every halo, we provide multiple proto-halo patches that include different amounts of the halo surrounding. The application combines these patches with the initial conditions of the simulation to a \music \citep{Hahn2011} configuration file that can be used to generate the initial conditions for various simulation codes.
    \item \emph{Referencing initial conditions.} Users can create references to halos and collections of halos that they used in their works, allowing the community to easily reproduce results and to compare different simulation codes and physics implementations. For each such reference, a tag is generated that can be included in research articles, allowing the community to access the zoom initial conditions using the \cosmicweb application or directly through a plugin distributed with \music .
\end{itemize}

This paper accompanies the release of \cosmicweb \footnote{\cosmicweb URL: \url{\cosmicweburl}}. and documents the simulations and pre-computed initial conditions that are available at release and the architecture and extendability of \cosmicweb. 
The structure is as follows: 
In \autoref{sec:sim}, we introduce the simulations that are included in the \cosmicweb application at release and give an overview of the available halo, galaxy and Lagrangian ellipsoid selectors which can be used to search the simulation databases. 
In \autoref{sec:proto-halos}, we discuss the computation of proto-halo patches from simulation data, in particular our choice of minimum bounding ellipsoids to describe Lagrangian proto-halos. We also present an analysis of the proto-halos provided on \cosmicweb by measuring their size and efficiency distributions depending on halo and environmental properties and the choice of traceback-radius.
In \autoref{sec:cosmicweb}, we present the individual components of \cosmicweb and how its interface can be used to find, download, and publish initial conditions for zoom simulations. 
In \autoref{sec:example_simulations}, we show two proof-of-concept zoom simulations to highlight the capabilities of \cosmicweb. 
We conclude in \autoref{sec:conclusions}. 
More details on the requirements of connecting additional simulations to \cosmicweb can be found in \autoref{sec:api_documentation}.


\section{Cosmological simulations and merger trees}\label{sec:sim}
The \cosmicweb platform has been designed in a modular way to allow for the integration of data from diverse projects hosted at different locations (also see \autoref{subsec:arch}).
Therefore, more simulations can easily be added in the future by us and the community. Here, we will briefly summarize the simulations that are available at the date of release. 

\begin{table*}
  \begin{threeparttable}
    \caption{List of simulations available at the time of release of \cosmicweb. The simulation parameters include the box length $L$ of the simulation volume, the cosmological parameters that were used, dark matter mass resolution $m_\mathrm{DM}$, gravitational force resolution $\epsilon$, number of snapshots, the structure finder that was used, and the minimum number of particles within the traceback-radius that we require to compute the minimum bounding ellipsoids, \(N^e_{\min}\).\label{tab:simulations}}
    \begin{tabularx}{\linewidth}{@{}llrXlrcccl@{}}
      \toprule
       & & $L$    & cosmology & $m_\mathrm{DM}$ & $\epsilon_\mathrm{com}$ & \(N^e_{\min}\) & \# snapshots & halo finder & G$^\dagger$\\
       & & \hMpc  &           & \hMsun          & \hkpc                   &                & [redshift range] &            & \\
      \midrule
      \multirow{5}{*}{\rotatebox[origin=c]{90}{primary}}
        & \agora             & 60    & Planck 2015   & \(1.21 \times 10^8\)  & 4   & 1000 & 101 [12.0 -- 0] & \rockstar     & \\
        & 150MPC             & 150   & Planck 2015   & \(2.70 \times 10^8\)  & 5   & 100  & 101 [12.0 -- 0] & \rockstar     & \\
        & 300MPC             & 300   & Planck 2015   & \(2.14 \times 10^9\)  & 10  & 100  & 101 [12.0 -- 0] & \rockstar     & \\
        & \rhapsody          & 1000  & WMAP 7/9      & \(6.46 \times 10^{10}\) & 50  & 1000 & 101 [12.0 -- 0] & \rockstar     & \\
        & \rhapsodypc        & 1000  & Planck 2015   & \(7.99 \times 10^{10}\) & 50  & 1000 & 101 [12.0 -- 0] & \rockstar     & \\
      \midrule
      \multirow{6}{*}{\rotatebox[origin=c]{90}{EAGLE}}
        & Ref-L0100N1504     & $100h$ & Planck 2014   & \(9.7 \times 10^6\,h\)  & $2.66h$ & 1000 & 29 [20.3 -- 0] & \textsc{fof}/\textsc{subfind} & \checkmark \\
        & Ref-L0025N0376     & $25h$  & Planck 2014   & \(9.7 \times 10^6\,h\)  & $2.66h$ & 1000 & 29 [20.3 -- 0] & \textsc{fof}/\textsc{subfind} & \checkmark \\
        & Ref-L0025N0752     & $25h$  & Planck 2014   & \(1.21 \times 10^6\,h\) & $1.33h$ & 1000 & 29 [20.3 -- 0] & \textsc{fof}/\textsc{subfind} & \checkmark \\
        & DMONLY-L0100N1504  & $100h$ & Planck 2014   & \(1.15 \times 10^7\,h\) & $2.66h$ & 1000 & 29 [20.3 -- 0] & \textsc{fof}/\textsc{subfind} & \\
        & DMONLY-L0025N0376  & $25h$  & Planck 2014   & \(1.15 \times 10^7\,h\) & $2.66h$ & 1000 & 29 [20.3 -- 0] & \textsc{fof}/\textsc{subfind} & \\
        & DMONLY-L0025N0752  & $25h$  & Planck 2014   & \(1.44 \times 10^6\,h\) & $1.33h$ & 1000 & 29 [20.3 -- 0] & \textsc{fof}/\textsc{subfind} & \\
      \bottomrule
    \end{tabularx}
    \vspace{1ex}
    \begin{tablenotes}
      \item[\(\dagger\)] This column indicates simulations that have been run with baryonic physics.
    \end{tablenotes}
  \end{threeparttable}
\end{table*}

\subsection{Gravity-only simulations}

Currently, we are locally hosting data from five gravity-only simulations. The parameters that were used for these simulations are summarized in \autoref{tab:simulations}. The simulations span cosmological volumes from $60^3$~$h^{-3}$Mpc$^3$ to $1$~$h^{-3}$Gpc$^3$ and use cosmological parameters conforming with the Planck 2015 \citep{Planck2015Parameters} and WMAP 7/9 \citep{Komatsu2011, Hinshaw2013} results.
We include two simulations which have been previously used for zoom simulations: the cosmological volumes of the AGORA comparison project \citep{Kim2013} and of the RHAPSODY cluster re-simulation project \citep{Wu2013a, Wu2013b, Hahn2017} which is originally based on the Carmen simulation of the LasDamas project \citep{McBride2009} -- we labeled the simulations \enquote{\agora} and \enquote{\rhapsody} for the sake of convenience. 
\rhapsody uses older WMAP 7/9  cosmological parameters. However, we have updated the parameters in the \rhapsodypc simulation to allow comparisons between the halo catalogs and zoom simulation results. 

Initial conditions for the full simulations were generated using \music \citep{Hahn2011} at redshift \(z=49\) (\rhapsody \& \rhapsodypc) and \(z=99\) (remaining simulations). Using the \textsc{L-gadget}-3 code -- a lean version of the \textsc{gadget}-3 code developed for the Millennium-XXL simulation \citep{Angulo2012} -- we evolved the simulations to redshift \(z=0\) and stored 100 snapshots between \(z=12\) and 0, logarithmically distributed in scale factor units. We adopted a fixed co-moving softening of $1/30$th of the linear particle separation for the ``primary'' runs. The numerical values of the force softening lengths for the ``primary'' and EAGLE simulations are listed in \autoref{tab:simulations}.

Each snapshot was processed with the \rockstar structure finder \citep{Behroozi2013}, which identifies halos and subhalos using a six-dimensional phase-space friend-of-friend algorithm and measures their properties, including spherical overdensity masses, radii, and concentrations, spin parameters, and relaxedness quantities. 
Using \consistenttrees \citep{Behroozi2013b}, we group halos across snapshots to merger trees containing information about spatial hierarchy (halo substructure) as well as temporal hierarchy (descendant and progenitor halos and merger events). 
In addition, we compute environmental parameters such as the normalized distance to the closest more massive neighbor \(D_{1,1}\) (cf. \autoref{tab:features}) directly from the halo catalog.

For every host halo in each snapshot, we compute the Lagrangian minimum bounding ellipsoid at $1, 2, 4$, and $10$~$R_\mathrm{vir}$, if the number of particles within that radius exceeds the threshold listed in \autoref{tab:simulations}. We will discuss the definition, construction, and statistical distribution of these proto-halo patches in more detail in \autoref{sec:proto-halos} and \autoref{sec:efficiency_distribution}.

The halo catalogs including the merger-tree information and the Lagrangian proto-halos are stored in a relational database. To reduce latency and speed-up queries, the data has been partitioned and indexed on the most commonly used columns. More details on the storage of and interface to the data can be found in \autoref{sec:local_api}.

\subsection{The \eagle database}
Another set of halo catalogs and merger trees originates from simulations of the \emph{Evolution and Assembly of GaLaxies and their Environments} simulation suite \citep[\eagle; ][]{Schaye2015, Crain2015, McAlpine2016} and is hosted by the Virgo consortium\footnote{\url{http://www.virgo.dur.ac.uk/}}. In particular, we provide access to the \texttt{Ref-L0025N0752}, \texttt{Ref-L0025N0376}, and \texttt{Ref-L0100N1504} simulation that were run with full gravity, hydrodynamics and subgrid modelling, and their \texttt{L0025N0752}, \texttt{L0025N0376}, and \texttt{L0100N1504} gravity-only counterparts. With the addition of simulations including full baryonic physics, \cosmicweb users can also use galaxy properties to constrain their search for zoom simulation targets.

The phases for the initial conditions for the \eagle simulations were extracted from regions within the large \panphasia multi-scale white noise field \citep{Jenkins2013}. The phases for the \eagle volumes are published in \citet{Schaye2015}.
The updated version of \music\footnote{The updated version 2 of \music is available from\\ \url{https://github.com/cosmo-sims/MUSIC}} includes the \panphasia white noise field, allowing us to create zoom simulations in the \eagle volumes from a unified interface.

The \eagle simulations were run with a modified version of the \textsc{gadget}-3 code with a full gravity and hydrodynamics treatment, including a large number of subgrid modules accounting for physical processes below the resolution scale, such as radiative cooling and heating, star formation and evolution, metal enrichment, and feedback from supernovae and supermassive black holes \citep{Crain2015}.
Each simulation contains 29 snapshots distributed between $z=20$ and $z=0$. In the snapshots, halos are identified using a FoF algorithm with linking length $b=0.2$ and a spherical over-density algorithm. Baryonic particles are then assigned to the FoF groups and all particle species within the group are further processed with the \textsc{subfind} algorithm \citep{Springel2001, Dolag2009} to separate self-bound structures and identify galaxies. The subhalo catalogs are linked across time by determining the descendant halos using \textsc{D-Trees} \citep{Jiang2014, Qu2017}, and the main progenitors defined by the largest \enquote{branch mass} \citep{Qu2017}.

The host halos for the Auriga Project galaxy formation simulations  \citep{Grand2017,Grand2024} were selected from the Eagle 100~Mpc side-length box. Table A1 in \citet{Grand2017} matches the host halos of the Auriga sample with their corresponding halo IDs in the Eagle database. In \autoref{sec:example_simulations}, we recreate the host halo of the Auriga 6 galaxy and compare the redshift zero dark matter halo properties against an original Auriga project dark matter simulation of that same halo.


\section{Proto-halo patches}\label{sec:proto-halos}

\begin{figure*}[ht]
    \centering
    \includegraphics[width=\textwidth]{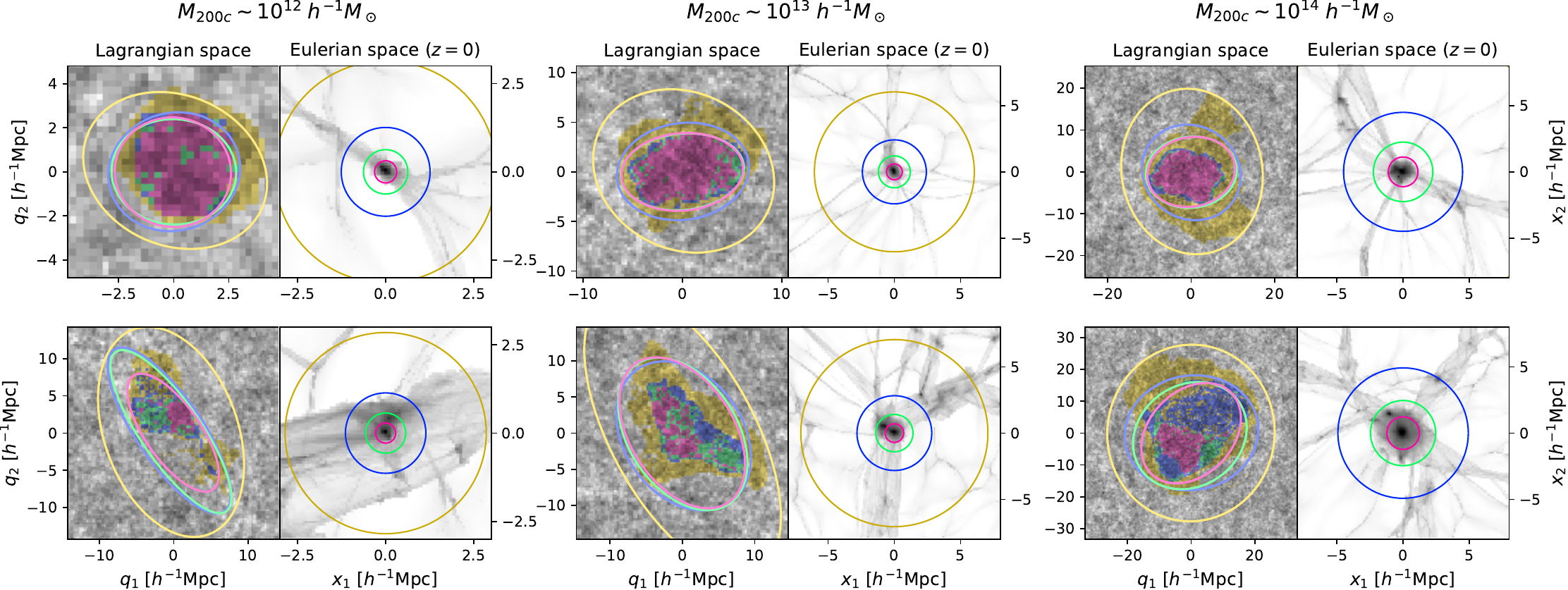}
    \caption{Slices through six halos at redshift \(z=0\) (right panels) and their proto-halo patches (left panels). The halos are selected from the 300MPC simulation, centered around three halo mass values \(\smash{M_{200,c} \sim 10^{12}, 10^{13}}\), and \(\smash{10^{14}}\)~\hMsun from isolated environments (top) and more populated regions (bottom). The circles in Eulerian space correspond to traceback-radii of 1 (magenta), 2 (green), 4 (blue), and 10 \(R_\mathrm{vir}\) (yellow), respectively. Particles within \(R_\mathrm{tb}\) are traced back to Lagrangian space and colored correspondingly. The ellipses represent slices through the minimum bounding ellipsoids and the background shading illustrate the density field (darker colors correspond to higher density).}\label{fig:protohalo}
\end{figure*}

In this section, we present an overview of the zoom regions available in \cosmicweb. Specifically, we define Lagrangian proto-halos and explain the rationale behind our choice of minimum bounding ellipsoids to describe these volumes.
We also present an analysis of the ellipsoids' mass and environment dependent efficiency, i.e. how well the ellipsoids are capturing the proto-halo shape. We measure the mass growth of the ellipsoids depending on the size of the zoom region and quantify the selection bias that is introduced if constraints on the efficiency are imposed.

Generally, a cosmological $N$-body simulation maps a fluid element, or $N$-body particle, from its initial unperturbed `Lagrangian' position $\bm{q}$ along a trajectory to a position $\bm{x}(\bm{q},z)$ in the evolved universe at redshift $z$. By definition thus, the Lagrangian coordinate $\bm{q}$ serves as a unique label of each particle over the course of a simulation. Tracing back the set of all particles $\mathcal{P}(\mathcal{H})=\{\bm{q}_i\}_{\bm{x}(\bm{q},z)\in\mathcal{H}}$ making up a halo $\mathcal{H}$ yields the proto-halo patch $\mathcal{P}$, i.e. the region of primordial Universe that will collapse to form this halo. In general, the set $\mathcal{P}$ will have irregular shape, and potentially even holes. 

Since the definition of a halo is to some degree arbitrary (see e.g. the discussion in \citet{White2001}) and, depending on the application, one in many cases is not only interested in a specific halo but also its surroundings, we allow for a range of traceback-radii centered at halo locations. 
This traceback-radius \(R_\mathrm{tb}\) of the halo, is chosen as a multiple of the halo radius defined by a spherical overdensity criterion.

Choosing a larger traceback-radius increases the computational cost of the simulation (although not proportionally to the volume, as the density decreases with the distance to the halo), but reduces the probability of contamination by low-resolution particles during the evolution of the halo.
A larger zoom region also allows to resolve the immediate environment of the halo in high resolution, which is desirable in many simulations. 
However, a too large traceback radius potentially includes surrounding halos, in particular if the targeted halo is located in a populated environment, resulting in a computationally more expensive simulation.

\autoref{fig:protohalo} illustrates the relation between proto-halos in Lagrangian space and the resulting halo at redshift \(z=0\). We choose six halos from the 300MPC simulation, two halos each around \(M_{200,c} \sim 10^{12}, 10^{13}\), and \(10^{14}\)~\hMsun, with one from an isolated and one from a more populated environment.  The traceback radii at 1, 2, 4, and 10 \(R_\mathrm{vir}\) are shown as circles in Eulerian space. All particles within the circles are traced back to the initial conditions and colored with the color corresponding to \(R_\mathrm{tb}\). We note that while in isolated environments, the proto-halo patches are compact and do not grow significantly with increasing traceback radius (with the exception of \(R_\mathrm{tb}=10R_\mathrm{vir}\)), they are less spherical, more scattered and grow significantly at larger radii in populated regions.

Once the particles within the traceback radius are identified, different techniques exist to describe the enclosing Lagrangian volume, such as a rectangular bounding box, minimum bounding ellipsoids, and convex hulls.  A comparison of some commonly used methods can be found in \citet{Onorbe2013}. Note that the refinement regions are generally chosen to be convex as concave patches and holes within the volume could contaminate the zoom region with low resolution particles and impact the accuracy of the simulation.
To measure the quality and computational cost of a proto-halo patch, we can define the efficiency as the ratio of the mass of the particles enclosed within \(R_\mathrm{tb}\) to the mass enclosed by the Lagrangian volume description, i.e.
\begin{equation}\label{eq:efficiency}
    \mathcal{E} = \frac{M_\mathrm{particles}}{M_\mathrm{ellipsoids}}.
\end{equation}
In general, the efficiency increases with the complexity of the boundary description, e.g the rectangular bounding box requiring 6 coordinates is less efficient than the minimum bounding ellipsoid (9 parameters), which in turn is less efficient than the convex hull described by all the particles on the surface \citep{Onorbe2013}. As we will show in \autoref{sec:efficiency_distribution}, the variance of the efficiency is mass dependent and increases for low-mass halos due to environmental effects, as most of the low-mass halos evolve in clustered environments with strong tidal fields.
In contrast, more massive halos form at the nodes of the cosmic web where mass accretion occurs more isotropically, leading to a more well-behaved, spherical proto-halo.

The efficiency parameter is a measure of how well the abstracted Lagrangian volume describes the proto-halo patch and determines the fraction of computational resources that is focused on the volume of interest within the traceback-radius. It is therefore desirable to create proto-halo patches with high \(\mathcal{E}\) for zoom simulations. 
The efficiency parameter naturally depends on the environment and formation history of the halo, which is why one has to be careful to not create a biased sample when selecting halos by the efficiency parameters of their Lagrangian minimum bounding ellipsoid. We will discuss this bias more quantitatively in \autoref{sec:efficiency_bias}. 

Generally, the ellipsoidal proto-halo patches provided in \cosmicweb are sufficient for many applications. However, for extremely high resolution simulations, users may want to further refine the boundary of the high resolution region in order to optimize computational resources. This can be easily achieved by first running a medium-resolution zoom simulation using the ellipsoidal patches in \cosmicweb and then calculating a more efficient boundary of the Lagrangian region using for example a convex hull algorithm.

\subsection{Minimal bounding ellipsoids in \cosmicweb}\label{sec:ellipsoids}
For \cosmicweb, we provide proto-halo regions in form of minimal bounding ellipsoids. This choice is motivated by the impracticality of storing convex hulls for all halos across all snapshots and multiple traceback-radii.
Minimal bounding ellipsoids offer a robust alternative while still capturing the overall shape of typical proto-halo patches at manageable data storage requirements. They furthermore provide a conservative choice in the sense that they are guaranteed to fill in potential holes in the Lagrangian patch.

An ellipsoid is described by its center $\bm{q}_c$ and its $3\times3$ shape matrix $\mathrm{A}$. The symmetric, positive definite  matrix $\mathrm{A}$ can be normalized such that coordinates $\bm{q}$ that are within the ellipsoid\footnote{A matrix is positive definite if one can write $\mathrm{A}=\mathrm{B}^\top \mathrm{B}$. The ellipsoid equation can thus also be written as that for a ball $\|\bm{p}\|^2<1$ where $\bm{p}:=\mathrm{B}(\bm{q}-\bm{q}_c)$ represents the inverse deformation of the ellipsoid to the sphere so that the eigenvalues of $\mathrm{B}$ are just the inverse principal axis lengths.} satisfy
\begin{equation}\label{eq:ellipsoid}
    (\bm{q} - \bm{q}_c)^\top \mathrm{A} \, (\bm{q} - \bm{q}_c) < 1.
\end{equation}
We outline the computed minimum bounding ellipsoids for the selected halos in \autoref{fig:protohalo}. Note that the illustration is only a 2-dimensional slice through the center of the ellipsoid, hence the ellipsoids appear to be a bad fit in some cases since the 3-dimensional structure is not visible. 
Nevertheless, we clearly notice that halos in isolated environments tend to have more ``efficient'' ellipsoids, with a higher fraction of particles within the proto-halo collapsing into the traceback-radius. 
We discuss the ellipsoid efficiency in a more statistical analysis in \autoref{sec:efficiency_distribution}.
Also note that some of the Lagrangian patches at low traceback radii contain holes, in particular for halos in more populated regions (bottom panels). These holes originate from particles that have previously traversed the halo and are now outside the overdensity radius. Sampling those holes with coarse particles would could lead to artifacts as they traverse the halo; however, using a minimum bounding ellipsoids as the refinement region will ensure that those particles that pass through the halo are also sampled in high resolution.

For the simulations provided by the \cosmicweb team (labeled ``primary'' in \autoref{tab:simulations}), we provide minimal bounding ellipsoids for free halos (i.e., no subhalos) at every snapshot and for four different traceback-radii \(R_\mathrm{tb} = 1, 2, 4\), and 10 \(R_\mathrm{vir}\) if the number of particles within \(R_\mathrm{tb}\) is larger than the threshold \(N^e_{\min}\). 
The EAGLE simulations provide minimum bounding ellipsoids at \(R_\mathrm{tb} = 1, 2, 4\), and 8 \(R_{200,c}\) for all FoF groups at each snapshot.
The ellipsoids are computed using Khachiyan's algorithm \citep{Khachiyan1996, Kumar2005}, which iteratively optimizes the volume of the ellipsoid under the constraint that all particles are contained.
We account for periodic boundary conditions by moving all particles to the same local coordinate system centered at the proto-halo center of mass. This requires that no axis of the proto-halo patch is longer than half the box size, which generally should not occur and for which a zoom-simulation would provide only marginal benefits.

Note that ellipsoids derived from small traceback radii, in particular from lower mass halos that are not well-sampled by particles, can cause contamination of the zoom region by coarse-resolution particles, leading to potential numerical artifacts. We recommend to choose ellipsoids from larger trace-back radii or to run an intermediate zoom simulation to refine the zoom region in this case.

\subsection{Efficiency distribution of minimum bounding ellipsoids}\label{sec:efficiency_distribution}

\begin{figure*}[t]
    \centering
    \includegraphics[width=\textwidth]{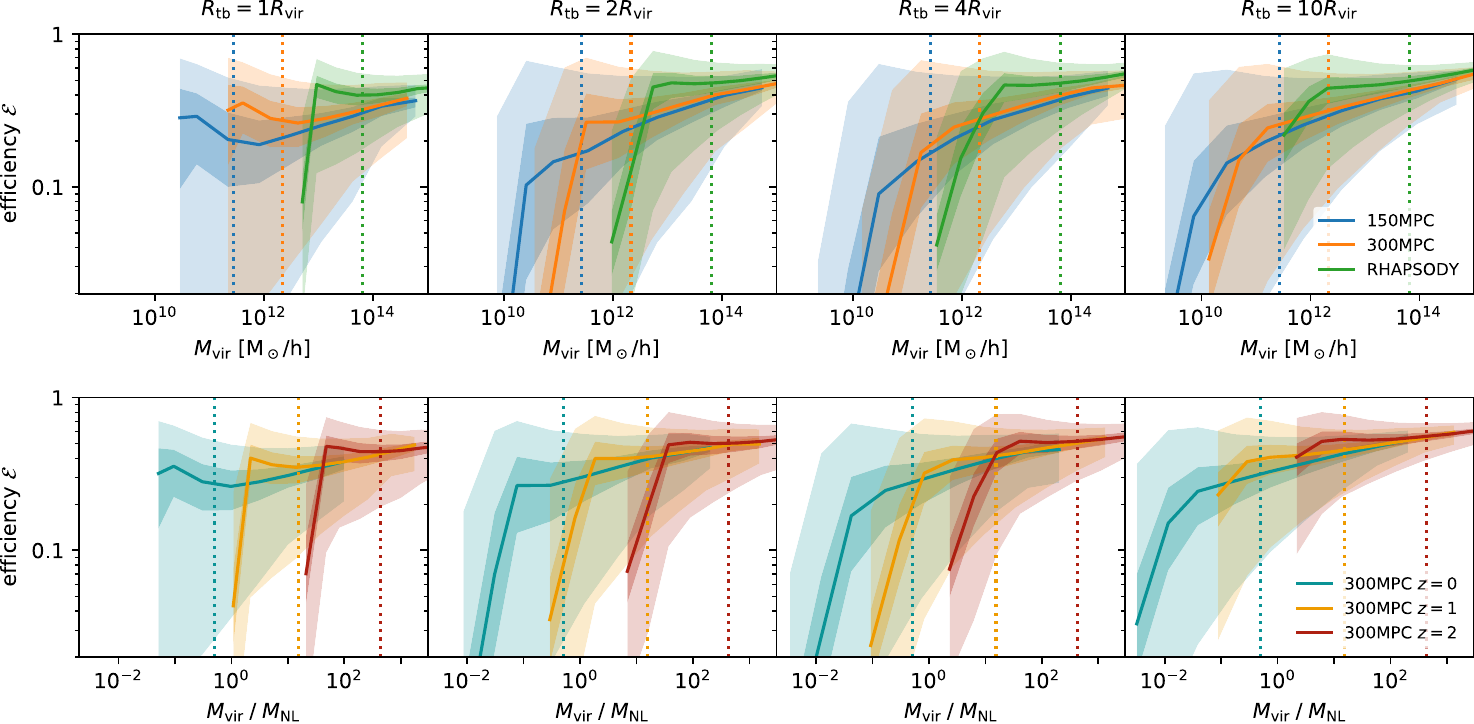}
    \caption{\textbf{Top:} Efficiency of the Lagrangian minimum bounding ellipsoids as a function of the halo mass (\(M_\mathrm{vir}\)). Shown is the median efficiency and the 66\% and 90\% contours. The minimal bounding ellipsoids are computed from particles within \(R_\mathrm{tb}=1, 2, 4\), and \(10 R_\mathrm{vir}\) of free halos in the 150MPC, 300MPC, and the \rhapsody simulation at redshift \(z=0\). The dashed vertical lines show the mass of 1000 particles for estimating the convergence threshold.
    \textbf{Bottom:} Efficiency of the Lagrangian minimum bounding ellipsoids from halos in the 300MPC simulation as a function of the halo mass (\(M_\mathrm{vir}\)) normalized by the nonlinear mass \(M_\star(z)\) at redshifts \(z=0, 1\), and 2.}\label{fig:eff}
\end{figure*}

\begin{figure*}[t]
    \centering
    \includegraphics[width=\textwidth]{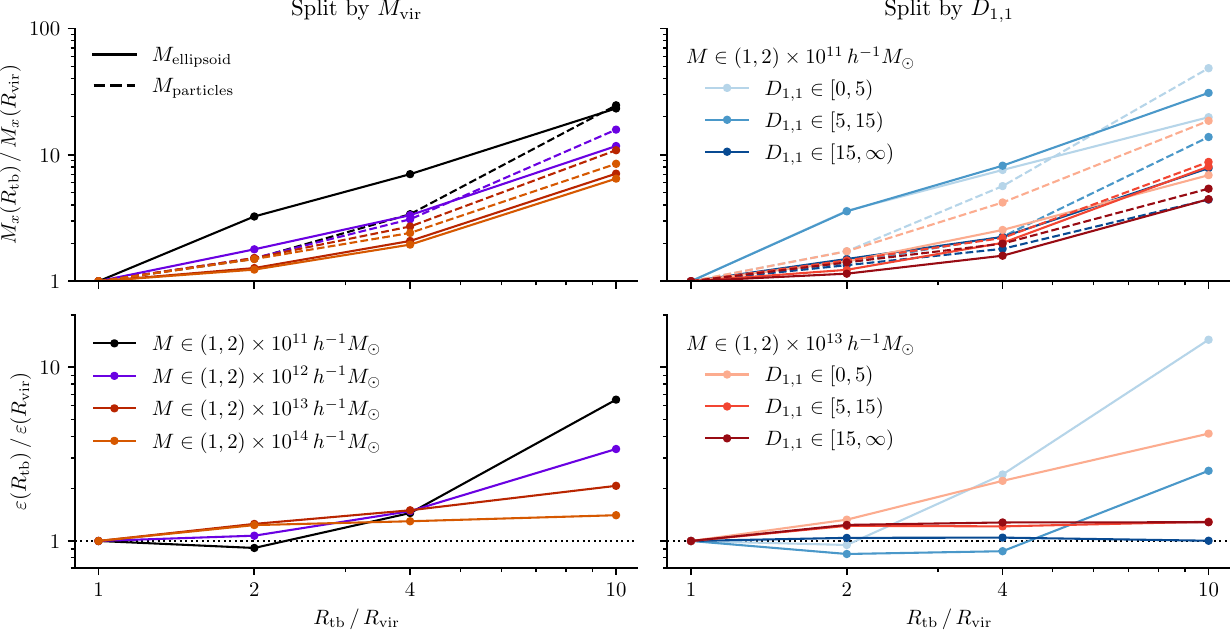}
    \caption{Relative change of mass of Lagrangian proto-halo patches (top) and the efficiency parameter (bottom) with increasing \(R_\mathrm{tb}\) for halos in the mass bins detailed in \autoref{tab:protohalo_massbins}.  For the mass increase, we plot the change in  \(M_\mathrm{ellipsoid}\) (dashed) and \(M_\mathrm{particles}\) (solid) separately. In addition, we subdivide two mass bins according to the environmental parameter $D_{1,1}$, shown on the right.}\label{fig:mass_rtb_split}
\end{figure*}

While high mass halos generally originate from more spherical patches \citep{Bardeen1986, Rossi2011}, lower mass halos in regions with strong tidal fields can have more elongated and irregular proto-halos, leading to a less efficient fit of the minimal bounding ellipsoid. 
This mass-efficiency relation has been analyzed in \citet{Onorbe2013}, which found an increased mean inefficiency and a larger standard deviation for lower mass halos.

In \autoref{fig:eff} (top), we show the median efficiency (see the definition in \autoref{eq:efficiency}) of the 150MPC, 300MPC, and the \rhapsodypc simulation, and the variance depending on the halo mass and the traceback-radius. As halo mass increases, the median efficiency increases at all traceback-radii and its distribution becomes more coherent. At the low-mass end, not all ellipsoids have converged due to the low number of particles  used to compute the ellipsoid. This can be seen by the upturn in the median efficiency for \(R_\mathrm{tb}=1\) and 2 and the drop for the lowest masses. We highlight the simulation dependent 1000-particle mass threshold, above which the ellipsoids with \(R_\mathrm{tb}=1R_\mathrm{vir}\) seem to converge. 

As halos grow, more particles from the outskirts fall within the traceback-radius. In the lower panel of \autoref{fig:eff}, we show the evolution of the ellipsoid efficiency in the 300MPC simulation at \(z=0, 1\), and \(2\). To account for the mass evolution, we normalize the halo masses by the nonlinear mass \(M_\mathrm{NL}(z)\) defined by 
\begin{equation}
    \sigma(M_\mathrm{NL}, z) = D_+(z) \sigma(M_\mathrm{NL}, z=0) = \delta_c,
\end{equation}
where \(\sigma(M, z)\) is the mass variance within a tophat kernel with radius \(R=(3M (4 \pi \bar{\rho})^{-1})^{1/3}\), \(D_+(z)\) is the linear growth function, and \(\delta_c = 1.686\) is the critical overdensity for spherical collapse. After correcting for the linear growth, efficiency distributions overlap among the analyzed snapshots, except for the upturn and downturn for the non-converged low-mass halos.

\begin{table}
  \begin{threeparttable}
    \caption{Details of the mass bins used for \autoref{fig:mass_rtb_split}, including which simulation the halos are obtained from, number of host halos in that mass range, and corresponding number of particles. Note that the number of particles within the ellipsoid is larger than the number of bound particles within the virial radius, since not all particles are gravitationally bound and proto-halos are not perfectly ellipsoidal.\label{tab:protohalo_massbins}}
    \begin{tabularx}{\linewidth}{@{}l r X r@{}}
      \toprule
      mass range & \# halos & simulation & \# particles \\
      $M_\mathrm{vir}$ [\hMsun] & & & per halo \\
      \midrule
      $(1, 2) \times 10^{11}$ & 51407 & 150MPC & 370 -- 740 \\
      $(1, 2) \times 10^{12}$ & 52030 & 300MPC & 470 -- 930 \\
      $(1, 2) \times 10^{13}$ & 6449  & 300MPC & 4670 -- 9340 \\
      $(1, 2) \times 10^{14}$ & 17803 & \begin{minipage}[t]{3cm}\textsc{Rhapsody\_}\\\textsc{PlanckCosmo}\end{minipage} & 1250 -- 2500 \\
      \bottomrule
    \end{tabularx}
  \end{threeparttable}
\end{table}

\begin{figure*}[t]
    \includegraphics*[width=\textwidth]{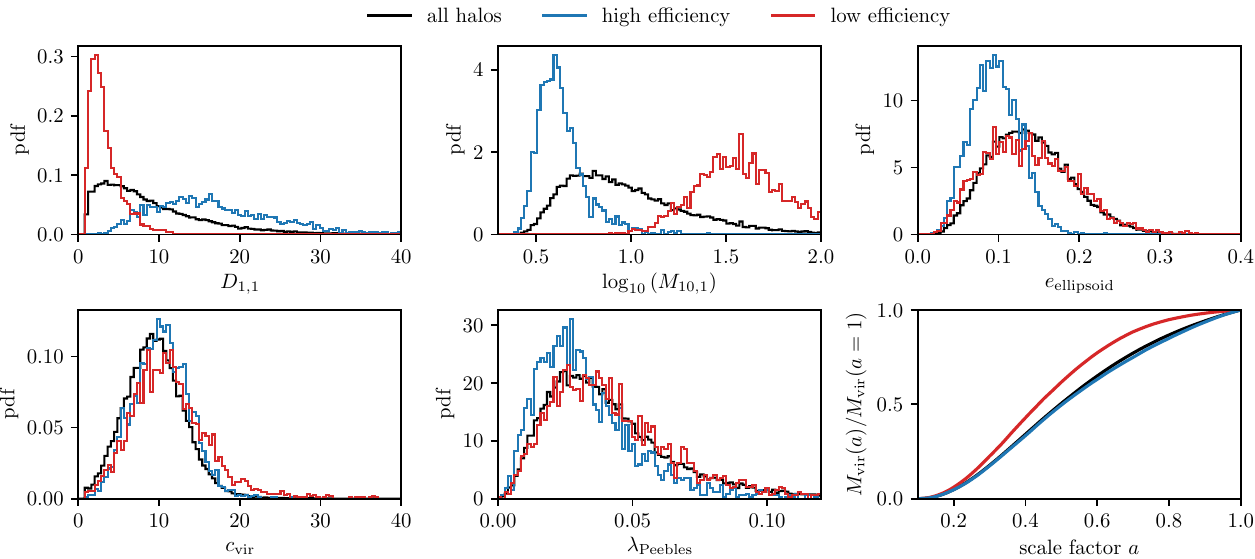}
    \caption{Distribution of some intrinsic and environmental halo properties of halos of the 300MPC simulation in the mass bin $M_\mathrm{vir} \in [2, 4] \times 10^{12}$\hMsun. 
    We separate the halos with proto-halo efficiency below the 10$^\mathrm{th}$ ($\varepsilon \simeq 0.013$) and above the 90$^\mathrm{th}$ ($\varepsilon \simeq 0.60$) percentile separately.
    Shown are the normalized distance to the closest more massive neighbor $D_{1,1}$, the relative mass within $10R_\mathrm{vir}$ $M_{10, 1}$, the ellipticity of the minimum bounding ellipsoid at $R_\mathrm{tb} = 2 R_\mathrm{vir}$, the concentration parameter $c_\mathrm{vir}$, the spin parameter $\lambda_\mathrm{Peebles}$, and the averaged mass history.}
    \label{fig:eff_parameter_histograms}  
\end{figure*}

Choosing a larger traceback-radius increases the volume of the minimal bounding ellipsoid. The exact growth of the Lagrangian ellipsoid depends on the mass and the environment of the halo.
To analyze the mass growth of the proto-halo patches at different traceback radii, we compare halos from four mass bins, detailed in \autoref{tab:protohalo_massbins}. 
Additionally, we split the $10^{11}$~\hMsun and $10^{13}$~\hMsun bins into three different environmental regimes, determined by the distance to the closest more massive halo $D_{1,1}$ (see \autoref{tab:features}).
The results are shown in \autoref{fig:mass_rtb_split}.

The mass of the particles within \(R_\mathrm{tb}\) as well as the minimal bounding ellipsoid, increases faster with increasing \(R_\mathrm{vir}\) for lower-mass halos and halos in more clustered environments.
For the lowest two mass bins, the relative mass growth of the Lagrangian ellipsoids is largest from 1 to 2 $R_\mathrm{vir}$ traceback-radii, whereas the proto-halos of the more massive halos grow faster at larger $R_\mathrm{tb}$.
In the case of low-mass halos, the minimum bounding ellipsoid grows faster than the mass within the traceback-radius, resulting in an efficiency drop at $R_\mathrm{tb} = 2 R_\mathrm{vir}$, as shown in the lower panel. The trend reverses for high-mass halos, where the number of particles within $R_\mathrm{tb}$ increases faster than the volume of the Lagrangian ellipsoid. For high-mass halos and for all halos at large $R_\mathrm{tb}$, the average efficiency therefore increases.

When splitting the mass bins according to the environmental parameter $D_{1,1}$, we find that the mass within $R_\mathrm{tb}$ rises more slowly for isolated halos than for halos in close proximity to a more massive neighbor. The presence of such a neighbor also leads to a faster growth of the  minimum bounding ellipsoid mass, lowering the mean efficiency parameter for the low-mass bin.

In summary, the efficiency of the minimum bounding ellipsoid, i.e., the ratio of the mass in the final target volume to the mass in the proto-halo patch, is markedly influenced by the halo mass and its environment. Higher mass halos and halos in isolated environments typically have a more spherical mass distribution that is better described by an ellipsoid compared to lower mass halos in dense, tidal regions. With the exception of the lowest mass bin, the numerical overhead, measured as the fraction of additional particles not part of the final target volume, decreases when the traceback-radius is increased.

\subsection{Biased target selection by efficiency constraints}\label{sec:efficiency_bias}

As discussed above, the environment clearly has a strong influence on the efficiency parameter. When choosing targets for zoom simulations, one might be inclined to select well-behaved proto-halos to minimize computational requirements. However, this approach potentially selects isolated objects, resulting in a biased sample within the chosen mass range. The environment is connected to other halo properties via a process commonly referred to as assembly bias \citep{Gao2005, Wechsler2006, Gao2007, Borzyszkowski2016}, which means these parameters will also be affected by the efficiency constraints. 

In order to measure these effects, we select halos with Lagrangian efficiencies below the 10$^\mathrm{th}$ and above the 90$^\mathrm{th}$ percentile from the $M_\mathrm{vir} \in [2, 4] \times 10^{12}$~\hMsun mass bin of the 300MPC simulation at $z=0$. We then compute the distributions of a selection of external and intrinsic parameters and compare it to the distribution of all halos in that mass bin.
The chosen parameters include the normalized distance to the closest more massive halo $D_{1,1}$, the relative mass increase between $1$ and $10 R_\mathrm{vir}$ ($M_{10,1}$), the concentration parameter $c_\mathrm{vir}$, the spin parameter $\lambda_\mathrm{Peebles}$ (cf. \autoref{tab:features}), and the ellipticity of the minimum bounding ellipsoid $e_\mathrm{ellipsoid} = (a_3 - a_1)/(2 \sum a_i)$, where $a_i = \lambda_i^{-1/2}$ are the semi-axes of the ellipsoid related to the eigenvalues $\lambda_1 \geq \lambda_2 \geq \lambda_3$ of the shape matrix (cf. \autoref{eq:ellipsoid}). This definition of ellipticity is the one of \cite{Bardeen1986}. The first two parameters measure the local clustering, while the following two are intrinsic properties of the halo. The ellipticity measures how spherical (and therefore well-behaved) the proto-halo patch is. Merger events and strong tidal fields are expected to distort the proto-halo patch, leading to a more elliptical and less efficient bounding ellipsoid.

In addition, we also measure the average mass growth history of the subsamples.
The results are shown in \autoref{fig:eff_parameter_histograms}. As shown above, the proto-halo efficiency and the local clustering are strongly correlated. Selecting only high-efficiency proto-halos results in a biased sample of preferentially isolated halos, with large $D_{1,1}$ and low mass increase $M_{10,1}$. Additionally, these proto-halo shapes are more spherical, indicating that the halos were not exposed to strong tidal fields during their evolution.

The effect on the halo concentration and the halo spin are weaker. Halos with high efficiency proto-halos tend to have lower spins, in agreement with predictions from e.g., Tidal-Torque Theory \citep{Doroshkevich:1970,White:1984}, in which the initial spin depends on the strength of the tidal tensor and its misalignment with respect to the proto-halo shape.
The mass evolution of high efficiency proto-halos follows the average of the entire mass-bin. However, low efficiency proto-halos accrete their mass early, with growth slowing down at earlier times compared to the the mass-bin average.


\section{Database architecture and web application}\label{sec:cosmicweb}
\begin{figure*}[t]
    \centering
    \includegraphics[width=0.75\textwidth]{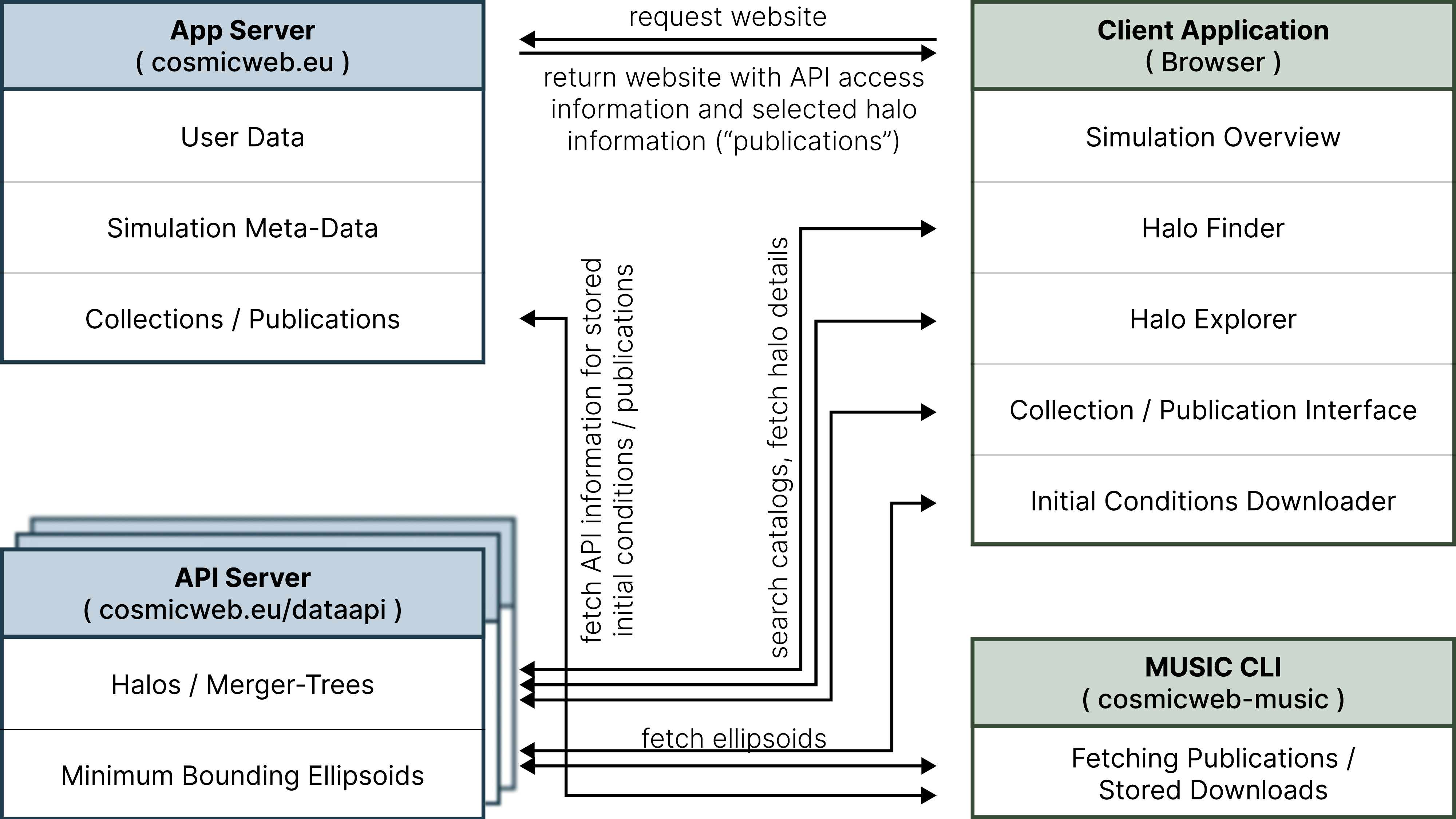}
    \caption{Sketch of the architecture of \cosmicweb. The server-side is divided into a main server, which hosts the web-server, manages user accounts, and stores user data and information, and several API servers that host the simulation data (halo catalogs and minimum bounding ellipsoids). Initial condition configurations for \music can be accessed via the website and the \cosmicwebmusic command-line tool.}\label{fig:arch}
\end{figure*}
In this section, we give an overview of the main components of the web application and discuss the implementation of the database that hosts the merger trees and proto-halo patches provided by the \cosmicweb team. As the implementation details may change in the future, this section reflects the state of the project at the date of release.

\subsection{Architecture}\label{subsec:arch}

Due to the diverse and potentially very large datasets, the \cosmicweb project is designed to be decentralized and modular. This allows us to host the merger trees and proto-halo patches on different servers than the application server, the user, and the publication data. We can therefore integrate existing databases to \cosmicweb without the need of duplicating data.

The \cosmicweb platform can be split into four main parts, visualized in \autoref{fig:arch}:
\begin{itemize}[itemsep=1pt]
    \item The data servers, which we will refer to as \emph{API servers}, providing access to the halo/galaxy catalogs and minimum bounding ellipsoids, 
    \item The application server hosting the web application and storing user data, simulation meta-data, and access information to the data servers,
    \item The web application running in the user's browser, and
    \item A command-line interface for the \music initial condition generator.
\end{itemize}

The main application server, hosted at \url{\cosmicweburl}, manages user accounts, stores collections and publications with references to the included halos, and provides connection information to the API servers for each supported simulation.
The server hosts the \cosmicweb website which features a JavaScript web application. This web-app loads additional data from the application and API servers, updating the webpage accordingly. Detailed information about the web application will be presented below.

Simulation metadata, including URLs, supported feature lists, and authentication tokens for connecting to the data servers, is dynamically loaded. This allows the web application to directly submit queries to the API servers and to download halo details and Lagrangian ellipsoids.

The API servers provide an interface between the actual datasets and the \cosmicweb application. This abstraction is essential to support heterogeneous databases through a common protocol. 

Finally, the \cosmicwebmusic command-line interface (CLI) is a convenient tool that allows users to download initial condition configuration files directly via the command line. By using the CLI, users can avoid the cumbersome process of downloading files locally and then re-uploading them to a headless compute cluster.

For the interaction between these four components to function seamlessly,  requests to and responses from the API servers must adhere to a specified protocol, which we further discuss in \autoref{sec:api_documentation}. More detailed and up-to-date documentation is also available directly on the \cosmicweb website\footnote{\url{\cosmicweburl/documentation/api}}.
How these requirements are best implemented on the API server side needs to be evaluated on a case-by-case basis, considering factors such as the available data, its storage, and the server infrastructure. We discuss the implementation of the API server used for the ``primary'' simulations in more detail at the end of this section.

\subsection{Web Interface}

\begin{figure*}[p]
    \centering
    \includegraphics[width=0.85\textwidth]{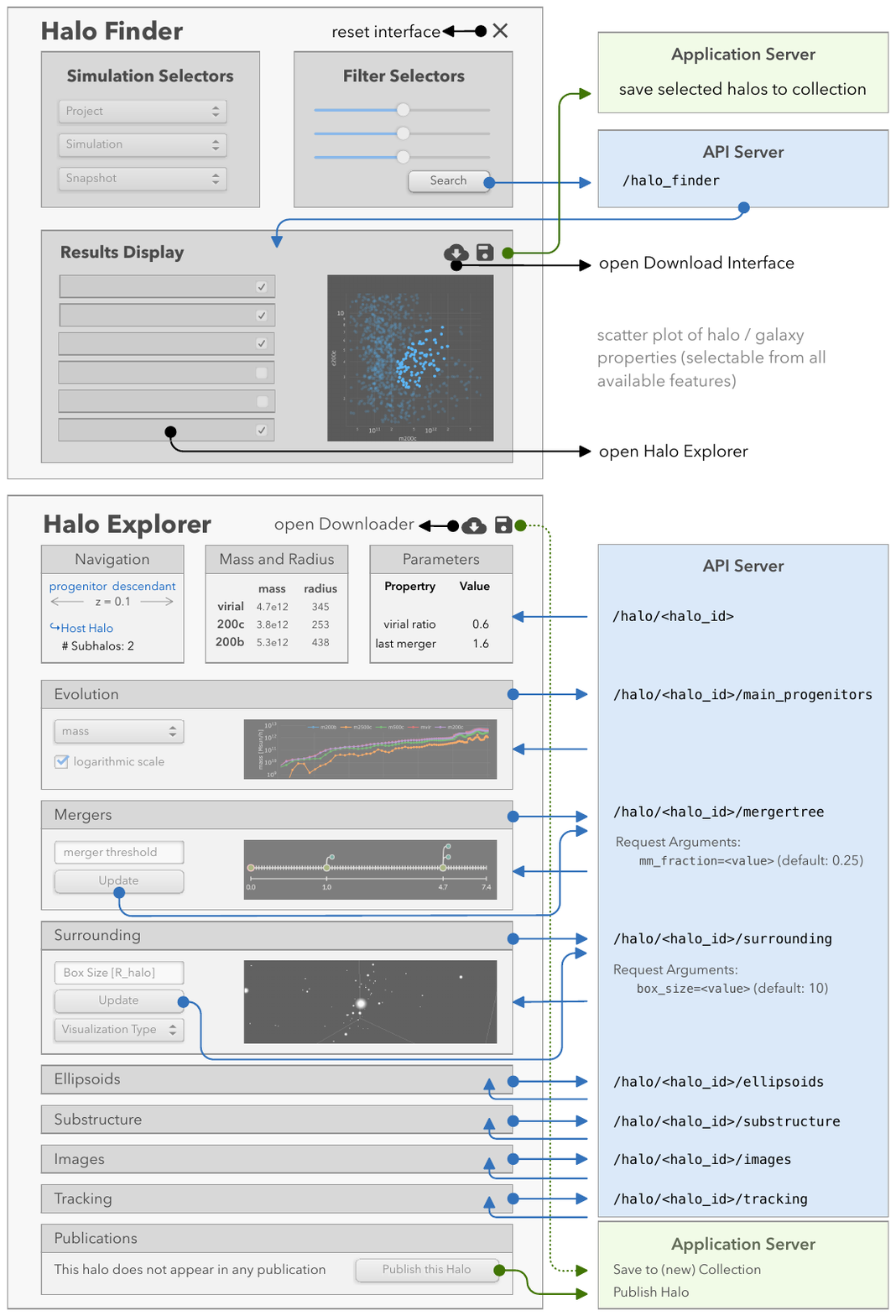}
    \caption{Illustration of the \emph{Halo Finder} and the \emph{Halo Explorer} window. Arrows indicate interactions with the API Servers (blue) and the main application server (green). Some screenshots of the \emph{Halo Explorer} visualizations are shown in \autoref{fig:screenshots}.}
    \label{fig:schematic_1}
\end{figure*}

\begin{figure*}[p]
    \centering
    \includegraphics[width=0.85\textwidth]{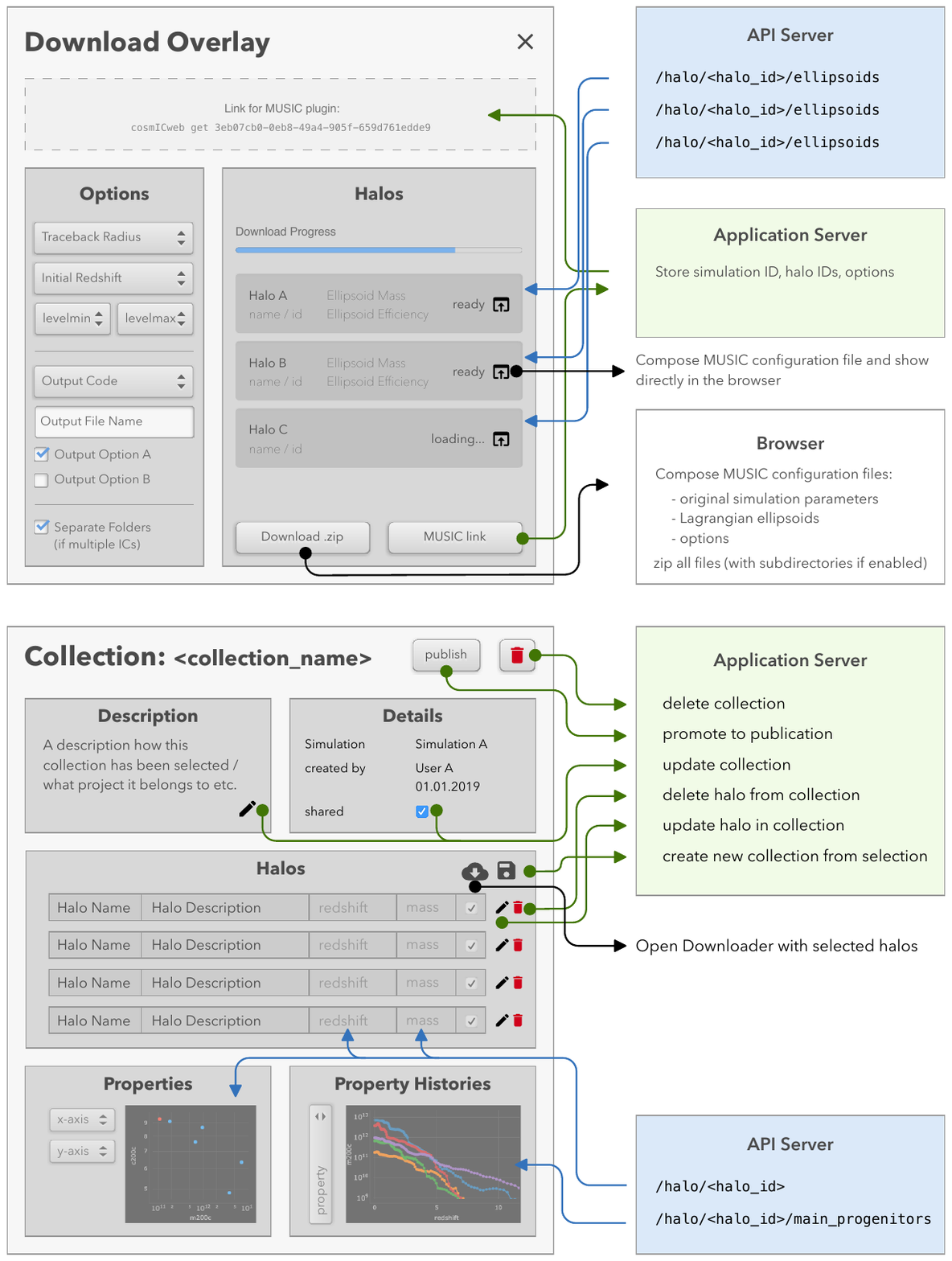}
    \caption{Illustration of the \emph{Download Interface} and the \emph{Collection Editor} window. Arrows indicate interactions with the API Servers (blue), the main application server (green), and with the client application itself (black).}
    \label{fig:schematic_2}
\end{figure*}

The \cosmicweb interface includes tools for searching halo and galaxy catalogs, analyzing and visualizing individual halos, configuring and downloading \music initial condition files, and creating, managing, and publishing collections of halos.
Here, we will briefly present each component and explain how they integrate into the zoom-simulation workflow.
Illustrations of the tools are shown in \autoref{fig:schematic_1} and \autoref{fig:schematic_2}, and we provide some screenshots of the visualization modules in \autoref{fig:screenshots}.

The \textbf{Halo Finder} is the main interface for searching halo and galaxy catalogs to identify target objects.
Users can select one of the available simulations and specify a snapshot or a range of redshifts. 
Limits for halo, galaxy, and ellipsoid properties can be set using either sliders or text input, with available filters depending on the selected simulation. \autoref{tab:features} provides an overview of the currently configured features.

\begin{table*}
  \begin{threeparttable}
  \caption{List of currently available selectors, grouped by halo properties, galaxy properties, and further parameters including properties of available proto-halo patches. [L] and [E] mark the features that are currently supported in the locally hosted ``primary'' simulations and the \eagle simulations.\label{tab:features}}
  \begin{tabularx}{\linewidth}{@{}l l c X l l@{}}
    \toprule
     & Parameter & Units & Description & L & E \\
    \midrule
    \multirow[c]{15}{*}{\rotatebox[origin=c]{90}{halo}} 
      & \(M_{200,c}\) / \(M_{500,c}\) / \(M_{2500,c}\) 
      & \hMsun 
      & Mass within \(\rho \ge \Delta \rho_c\) contrast 
      & \checkmark 
      & \checkmark \\
    & \(M_{200,b}\) / \(M_{500,b}\) / \(M_{2500,b}\) 
      & \hMsun 
      & Mass within \(\rho \ge \Delta \bar{\rho}\) contrast 
      & \checkmark\tablenotemark{*} 
      & \checkmark\tablenotemark{$\dagger$} \\
    & \(M_\mathrm{vir}\) 
      & \hMsun 
      & Mass within the virial boundary using the definition of \(\rho_\mathrm{vir}\) from \citet{Bryan1998} 
      & \checkmark 
      & \\
    & \(R_{\Delta c}\) / \(R_{\Delta b}\) / \(R_\mathrm{vir}\) 
      & \(h^{-1}\mathrm{kpc}\) 
      & Radius of the halo according to the mass definition 
      & \checkmark\tablenotemark{*} 
      & \checkmark\tablenotemark{$\dagger$} \\
    & \(c_{\Delta c}\) / \(c_{\Delta b}\) / \(c_\mathrm{vir}\) 
      & --- 
      & concentration parameter \(c_x = R_x / R_s\), where \(R_s\) is the scale radius of the NFW profile 
      & \checkmark\tablenotemark{*}\tablenotemark{$\diamond$} 
      & \\
    & \(\lambda_\mathrm{Peebles} = J \sqrt{|E|}\,/\,(GM_\mathrm{vir}^{5/2})\) 
      & --- 
      & Spin parameter (Peebles definition, \citet{Peebles1971})\tablenotemark{$\ddagger$} 
      & \checkmark 
      & \\
    & \(\lambda_\mathrm{Bullock} = J\,/\,\sqrt{2GM_\mathrm{vir}^3R_\mathrm{vir}}\) 
      & --- 
      & Spin parameter (Bullock definition, \citet{Bullock2001})\tablenotemark{$\ddagger$} 
      & \checkmark 
      & \\
    & \(f_\mathrm{sub}\) 
      & --- 
      & Substructure parameter \(\sum_i M_{\mathrm{sub}, i} / M_\mathrm{host}\) 
      & \checkmark 
      & \\
    & virial ratio 
      & --- 
      & Ratio between kinetic and potential energy \(2T / U\) 
      & \checkmark 
      & \\
    & \(x_\mathrm{offset}\) 
      & --- 
      & Normalized offset between center of mass and the most bound particle\newline
        \(x_\mathrm{offset} = \|\bm{x}_\mathrm{com} - \bm{x}_\mathrm{core}\|/R_h\) 
      & \checkmark 
      & \\
    & \(D_{1,1}\) 
      & --- 
      & Normalized distance to the closest more massive neighbor\newline
        \(D_{1,1} = \|\bm{x}_h - \bm{x}_n\|/R_h\) (\(n\): neighbor) 
      & \checkmark 
      & \checkmark \\
    & last major merger 
      & --- 
      & redshift of last major merger (with a mass ratio of 1:3 or greater) 
      & \checkmark 
      & \\
    \midrule
    \multirow[c]{4}{*}{\rotatebox[origin=c]{90}{galaxy}} 
      & \(M_\mathrm{dm}\) 
      & \hMsun 
      & dark matter mass in the central galaxy / subhalo 
      & 
      & \checkmark \\
    & \(M_\mathrm{star}\) 
      & \hMsun 
      & stellar mass in the central galaxy / subhalo 
      & 
      & \checkmark \\
    & \(M_\mathrm{gas}\) 
      & \hMsun 
      & gas mass in the central galaxy / subhalo 
      & 
      & \checkmark \\
    & \(M_\mathrm{BH}\) 
      & \hMsun 
      & black hole mass in the central galaxy / subhalo 
      & 
      & \checkmark \\
    \midrule
    \multirow[c]{4}{*}{\rotatebox[origin=c]{90}{others}} 
      & \(z\) 
      & --- 
      & redshift (range) 
      & \checkmark 
      & \\
    & \(n_\mathrm{part}\) 
      & --- 
      & Number of particles used to construct the minimum bounding ellipsoid 
      & \checkmark 
      & \\
    & \(\epsilon\) 
      & --- 
      & Efficiency of the Lagrangian minimum bounding ellipsoid\newline
        \(\epsilon = n_\mathrm{part} / V_\mathrm{ellipsoid}\) 
      & \checkmark 
      & \\
    \bottomrule
  \end{tabularx}
  \vspace{1ex}
  \begin{tablenotes}
      \setlength{\itemsep}{1ex}
      \item[*] The primary simulations currently do not support \(M_{500,b}\), \(M_{2500,b}\) and the corresponding radii and concentration parameter.
      \item[\(\diamond\)] The primary simulations use the scale factor \(R_s\) calculated by Rockstar from \(M_\mathrm{vir}\) and \(v_\mathrm{max}\) \citep{Behroozi2013}. Note that for concentrations other than \(c_\mathrm{vir}\), there is a systematic shift in the concentrations derived from this \(R_s^\mathrm{vir}\) compared to the \(R_s\) calculated from the corresponding overdensity definition directly, see \citet{Mansfield2020}.
      \item[\(\dagger\)] The \eagle simulations currently support \(M_\mathrm{200,b}\), \(M_\mathrm{200,c}\), \(M_\mathrm{500,c}\), \(M_{\mathrm{2500},c}\), and \(R_\mathrm{200,c}\).
      \item[\(\ddagger\)] \(J\) corresponds to the magnitude of the halo angular momentum and \(E\) to the total energy of the halo.
    \end{tablenotes}
  \end{threeparttable}
\end{table*}

Once a query is submitted, the API server returns a list of matching halos. The properties of these halos can be displayed in a table and visualized on  two-dimensional scatter plost. From the results panel, users can open halos in the Halo Explorer for a more detailed overview, group and store them as collections, and directly download  as initial condition files for resimulation.

The \textbf{Halo Explorer} lists and visualizes the properties of a single halo and allows users to navigate the substructure and merger tree. The available information varies depends on the simulation. We have implemented several visualizations, including a graphical schematic of the substructure, a merger tree with a customizable merger-ratio threshold, a graph showing the evolution of any halo property along the main-progenitor branch, a three-dimensional visualization of the halo surrounding using \textsc{WebGL}, a tracking tool for the halo position which can be used to configure movie tools in various simulation codes, and a graphical visualization of the Lagrangian minimum bounding ellipsoids at the available traceback-radii. Some screenshots are provided in \autoref{fig:screenshots}.
From the \emph{Halo Explorer}, users can directly download the \music configuration file for the selected halo and add the halo to new or existing collections for a later analysis or publication.

When querying halos in the \emph{Halo Finder}, numerous halos may meet the search criteria and be suitable for re-simulation. Narrowing down the selection requires a more thorough investigation of the candidates using the \emph{Halo Explorer}, or possibly running medium-resolution zoom simulations on a few targets. It is convenient to temporary store the candidates and share them with colleagues. Collections are designed to fulfill this need by providing an intuitive interface for grouping halos, with the capability to name and add a description to both the group and individual halos.

Collections are viewed and managed in the \textbf{Collection Editor}. This tool allows users to name individual halos and add short descriptions to highlight why the halos were chosen and what their characteristics are.
The collection interface includes visualizations to compare the properties of the halo sample and their evolution history along the main progenitor branches.

If the collection is being used in a published article, a permanent public URL to the collection can be created and included in the article\footnote{If the article references a single halo only, a publication link to this object can also be created directly from the \emph{Halo Explorer} without having to create a collection.}. The collection will then appear under the list of publications on \cosmicweb, along with the description and a link to the research article. 

Finally, the \textbf{Download} interface allows users to obtain the \music configuration files for one or multiple halos of the same simulation. 
The overlay includes options for the traceback radius of the minimum bounding ellipsoid and various \music parameters such as the starting redshift, the coarsest and finest particle resolution of the simulation, and the type of initial conditions file that will be generated. These options are added to the configuration file, facilitating the batch processing of multiple zoom simulations. Depending on the selected output format, additional options specific to that code are shown in the interface.

After specifying the options, the interface offers three different methods for retrieving the initial conditions configuration files: clicking on an individual entry to open a single configuration directly in the browser, downloading an archive containing \music files for all halos, or storing the configuration online to later download the configurations using the \cosmicwebmusic tool. This allows users to download the initial conditions directly on the server where the full initial conditions will be generated.

\subsection{API implementation for the \cosmicweb datasets}\label{sec:local_api}
The ``primary'' simulations provided by the \cosmicweb team support all optional features and visualizations, except for images (see \autoref{tab:api_endpoints}). The server thus needs to be able to efficiently access halo catalogs with merger-tree information and associated Lagrangian ellipsoids. 
To achieve this, we store the data in a SQL database, with halo-relationships, such as descendant halos, main progenitor halos, host halos, and main subhalos, expressed as foreign keys. 
Merger- and substructure trees can then be efficiently retrieved using recursive queries.
To minimize storage requirements, properties such as the fraction of mass in substructure, half mass time, and major mergers with custom mass ratios are computed on the fly.

Lagrangian ellipsoids for the \cosmicweb simulations have been obtained from particles within 1, 2, 4, and 10 $R_\mathrm{vir}$ of the halos.
We store the centers and shapes of these minimal bounding ellipsoids, along with the number of particles used to compute the ellipsoid and the resulting efficiency, in a separate table. This additional information helps identify and filter unconverged and inefficient ellipsoids. 

To parse incoming API requests to various endpoints, dispatch queries to the database, and format the response according to the API specifications, we use the \textsc{Flask} framework in combination with the \textsc{SQLAlchemy} library. 
Some complex and time-intensive queries, such as building merger trees or hierarchical trees, have been manually profiled and optimized to ensure reasonable execution times. To identify future bottlenecks in the API server, queries that take a disproportionate amount of time to process are logged on the server, allowing us to optimize the application as needed.

\section{Example Simulations}
\label{sec:example_simulations}
As a proof-of-concept, we present two zoom simulations originating from the \eagle RefL0100N1504 volume. In addition to demonstrating the capability of \cosmicweb, these examples also serve as a check of the \panphasia white-noise field integration into \music. One of the challenges to incorporate \panphasia is that (currently) \music requires power-of-two box resolutions due to the implementation of a multigrid Poisson solver to compute LPT terms \citep[see][]{Hahn2011}. This constraint requires a Fourier-space remapping of \panphasia fields. Specifically, we generate the \panphasia fields at the native resolution, e.g., $3\times 2^\ell$ where $\ell\in\mathbb{N}$ is some subdivision level, and then cut out the modes representable at the next smaller power of two (in the example $2^{\ell+1}$). This naturally introduces some small discrepancies and warrants an investigation how well we can reproduce existing zoom simulations with \music and its \panphasia integration.

Note that \music allows to specify a minimum refinement level for the entire simulation, $l_\mathrm{min}$, which sets the number and mass of the coarsest particles in the zoom simulation, and a minimum refinement level when generating the initial conditions, $l_\mathrm{min}^\mathrm{TF}$. The particle displacements and velocities are later averaged down to $l_\mathrm{min}$. We present simulations varying both. This tests the dependence on the coupling of small-scale modes into the high-res regions, i.e. the accuracy of the zoom approach itself as well as its implementation in \music, and the impact of a coarsely represented tidal field on the non-linear evolution during the simulation.

\subsection{Milky-Way sized halo}

\begin{figure}
    \centering
    \includegraphics[width=\linewidth]{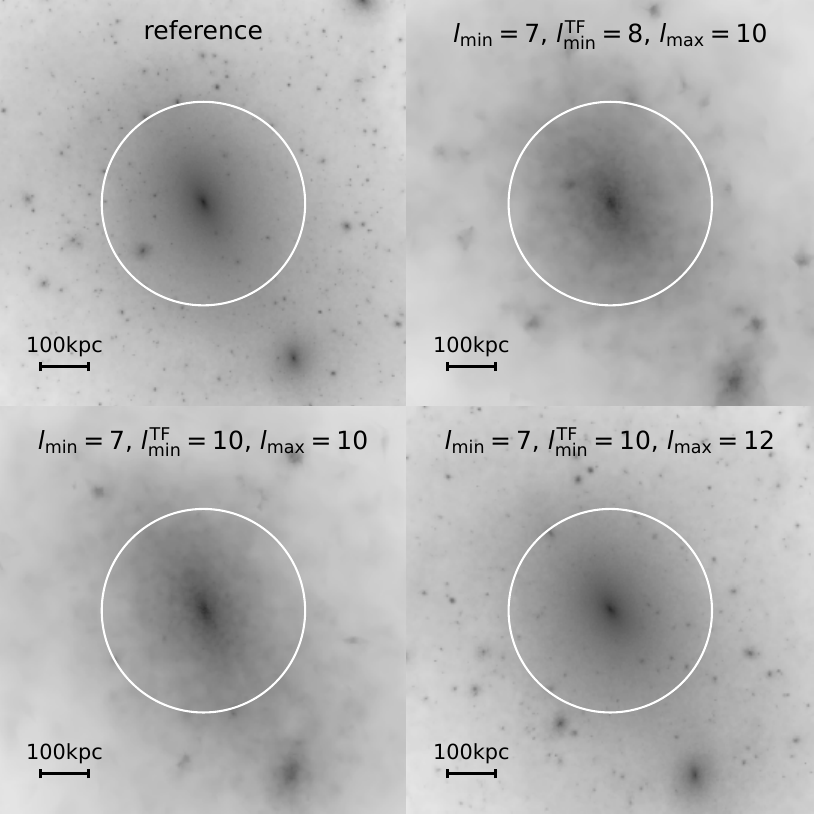}
    \caption{Visual comparison of the Au6 halo \citep{Grand2017} between the reference simulation and three zoom simulations (cf. \autoref{fig:auriga6_profiles}). We calculate the surface density using a Delaunay tessellation field estimator \citep{Rangel2016}. The white circle indicates $R_\mathrm{200,c}$ determined in the reference simulation.}
    \label{fig:auriga6_projection}
\end{figure}

\begin{figure}
    \centering
    \includegraphics[width=\linewidth]{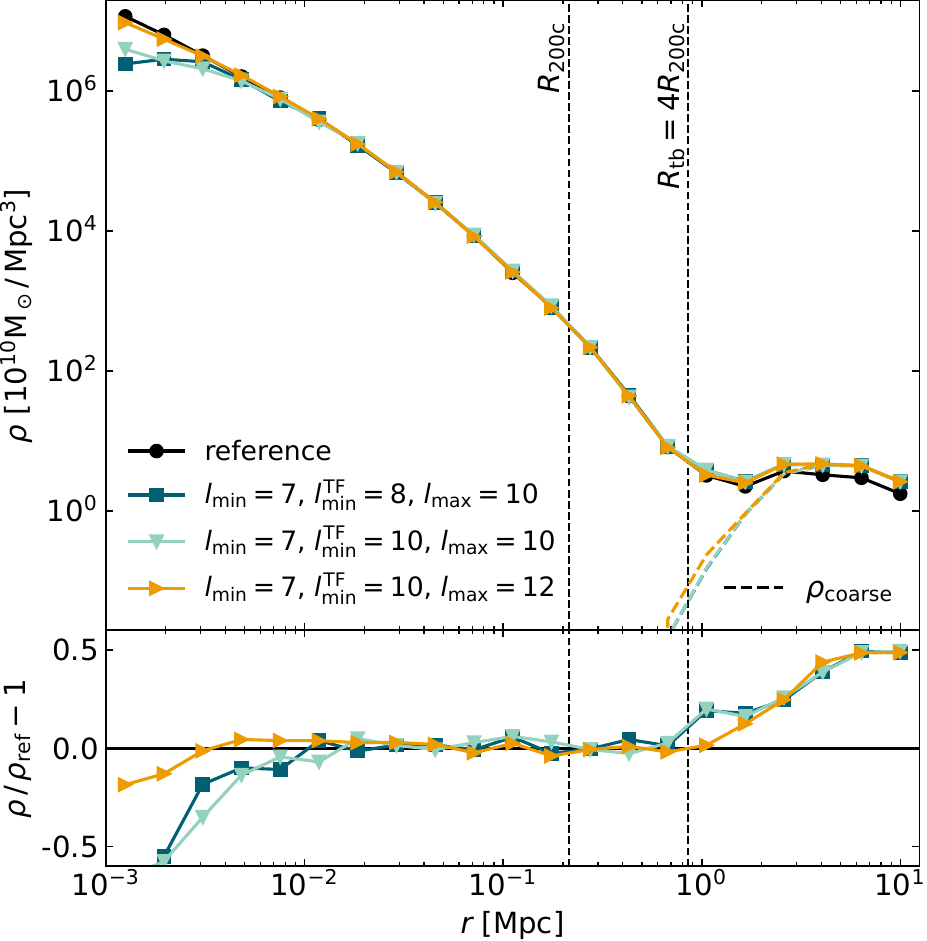}
    \caption{Density profile comparison for the Au6 halo \citep{Grand2017}. We compare three zoom simulations initialized with \music using proto-halo patches from \cosmicweb against the reference simulation by \citet{Grand2017} that has been initialized with \panphasia. The proto-halo patches have been calculated at a traceback radius of $4 R_\mathrm{200,c}$. In addition to the total matter density profiles (solid) we highlight the contribution of low resolution particles outside the refinement region separately (dashed). The lower panel shows the relative difference of the total matter density profile with respect to the reference solution.}
    \label{fig:auriga6_profiles}
\end{figure}

\begin{table}
  \begin{threeparttable}
    \caption{Global halo properties of the Milk-Way-sized Auriga~6 halo in the re-simulations at different resolution -- see also \autoref{fig:auriga6_projection} for images. Reference values have been re-measured from the original simulation \citep{Grand2024} with consistent \rockstar parameters and slightly differ from the values reported in \citet{Grand2024}.\label{tab:auriga_props}}
    \begin{tabularx}{\linewidth}{@{}X rrrr@{}}
      \toprule
       & (7,8,10) & (7,10,10) & (7,10,12) & Ref. \\
      \midrule
      $m_{p}$ [$10^5\mathrm{M}_\odot$] & 364 & 364 & 5.69 & 2.34 \\
      $\epsilon_{\rm comoving}$ [$\mathrm{kpc}$] & 3.26 & 3.26 & 0.813 & 0.739 \\
      \midrule
      $M_{200,c}$ [$10^{10}\mathrm{M}_\odot$]  & 105.3 & 108.1 & 105.2 & 105.4\\
      $M_{500,c}$ [$10^{10}\mathrm{M}_\odot$]  & 83.2  & 83.3  & 83.0  & 82.0\\ 
      $M_{2500,c}$ [$10^{10}\mathrm{M}_\odot$]  & 48.0  & 48.0  & 49.8  & 48.5\\ 
      $R_{200,c}$ [kpc]                   & 214.5 & 216.3 & 214.4 & 214.5\\
      $R_{s}$ [kpc]                       & 22.7  & 24.2  & 21.1  & 22.5\\
      $c_{200,c}$                         & 9.45  & 8.95  & 10.18 & 9.53\\
      \bottomrule
    \end{tabularx}
  \end{threeparttable}
\end{table}

The first example is the Au6 halo from the Auriga Project \citep{Grand2017,Grand2024}. This Milky Way mass halo has previously been studied to analyze formation mechanisms of disc galaxies. We perform gravity only zoom simulations at different levels of refinement of the Au6 Lagrangian patch, calculated at a traceback radius $R_\mathrm{tb} = 4 R_\mathrm{200,c}$ at redshift 0. The halo can be found in the publication section on \cosmicweb\footnote{Auriga 6: \url{https://cosmicweb.eu/publication/auriga-halos}}.

We re-simulated this halo at two different mass resolutions ($l_\mathrm{max}=10$ and 12). In addition, we once generated the initial noise field at the full resolution followed by averaging in the coarse regions ($l_\mathrm{min}^\mathrm{TF}=l_\mathrm{max}=10$ : (7,10,10)), and using the zoom technique ($l_\mathrm{min}^\mathrm{TF}$ $<$ $l_\mathrm{max}$: (7,8,10) and (7,10,12)).
The re-simulation was carried out using the \textsc{swift} $N$-body code \citep{Schaller2024}. The respective mass resolution of all runs along with the force softening are indicated in \autoref{tab:auriga_props}.
We compare these halos with the redshift zero output of a reference calculation, which is an  original Auriga level~4 gravity only simulation run with \textsc{gadget}. The initial conditions for the reference were created by the \icgen code \citep{Jenkins2010,Jenkins2013} using a $1536^3$ cell patch from the \panphasia field. The nearest smaller representable resolution in \music is $l_\mathrm{min}^\mathrm{TF}$=10 with $2^{10}=1024$.  We determined halo properties using the \rockstar halo finder \citep{Behroozi2013} and list the recovered main properties of the halos in \autoref{tab:auriga_props}, as well as show the projected mass distribution in \autoref{fig:auriga6_projection} and spherically averaged mass density profiles in \autoref{fig:auriga6_profiles}. 

Regarding global properties, we find that the halo mass $M_\mathrm{200,c}$ is recovered at the level of better than 1 per cent for two re-simulations, while the third, $(7,10,10)$, is somewhat more discrepant at $\sim 3$ per cent. The virial radius always agrees to better than one per cent. Properties of the inner mass distribution (concentration $c_{200,c}$ and mass $M_{2500,c}$) of the halos are slightly more discrepant but still consistent at the few per cent level. Note that the \rockstar measurement of the mass and other properties of the reference run is slightly discrepant from the values reported in \citet{Grand2024} using \textsc{subfind}.

The visual impression from \autoref{fig:auriga6_projection} is similar in the overall shape of the halo, also positions of some halos outside the virial radius are in excellent agreement. Interestingly, the most massive substructure is however at different locations in all of the runs --- likely a consequence of it being very sensitive to any perturbations of the initial conditions. Such sensitivities are expected since orbits in virialised halos can be chaotic (see e.g. \citet{Vogelsberger:2008,Sousbie:2016,Genel:2019} for some degree of evidence for the presence of chaotic mixing) implying that even small perturbations can have impact on fine-grained properties of halos, such as specifically the exact positions of substructures (see e.g. also \citet{Wang:2021}).

Finally, the density profiles shown in \autoref{fig:auriga6_profiles} show agreement at the few per cent level at all radii up to the radius affected by finite force and mass resolution and out to the maximum radius of the zoom region of $4R_{200,c}$. 
The spherically averaged density profile is well reproduced despite the actual positions of the subhalos differing between simulations.  This is possible simply because the mass associated with these substructures is small.

\subsection{Group-sized halo}

\begin{figure*}[ht]
    \centering
    \includegraphics[width=\linewidth]{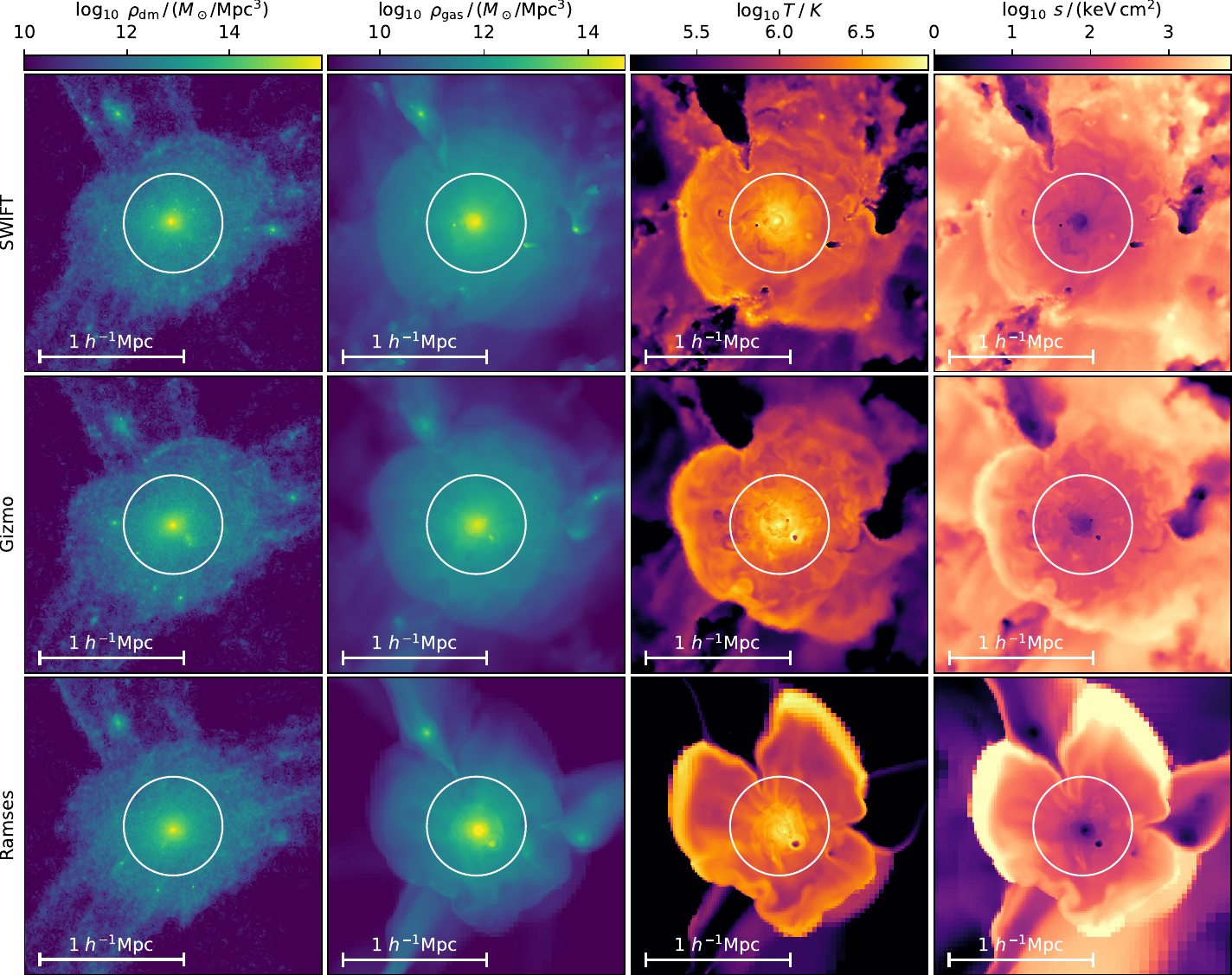}
    \caption{Comparison of the dark matter density and gas properties between three re-simulations of the halo described in \citet{Richings2021} using \textsc{swift} \citep{Schaller2024} (top), \textsc{gizmo} \citep{Hopkins2015} (middle), and \textsc{ramses} \citep{Teyssier2002} (bottom). The four columns show the dark matter density, gas density, gas temperature, and gas entropy in a slice perpendicular to the $z$-axis centered on the halo. The white circle indicates the $R_{200,c}$ spherical overdensity radius measured in the \textsc{gizmo} run. See main text for simulation details.}
    \label{fig:richings_bary4}
\end{figure*}

The second halo we resimulate is a group-mass halo that has been featured in a study quantifying the impact of baryonic physics on halo and substructure properties \citep{Richings2021}. This halo is also from the Eagle 100~Mpc volume and its GroupID from the Eagle database in the reference model is 129 in snapshot 26 ($z=0.18$)\footnote{Initial conditions on \cosmicweb can be accessed under \url{https://cosmicweb.eu/publication/group-sized-halo-richings-2021}}.

The initial conditions for the resimulation have been generated with \music using $l_\mathrm{min}=7$, $l_\mathrm{min}^\mathrm{TF}=10$, and $l_\mathrm{max}=11$, with the zoom region obtained from the $R_{tb}=8R{200,c}$ Lagrangian ellipsoid. 
For these resimulations, we follow a different strategy in that we demonstrate how \cosmicweb facilitates the reproducibility of cosmological simulation results across different simulation codes by leveraging the plethora of output plugins available for the \music code. Specifically, we show resimulations, at fixed resolution of the initial conditions, using the \textsc{ramses} \citep{Teyssier2002} code (Eulerian adaptive mesh finite volume hydrodynamics and adaptive mesh particle mesh gravity solver), the \textsc{gizmo} \citep{Hopkins2015} code (Mesh-free Lagrangian finite volume hydrodynamics with a tree-particle~mesh gravity solver), and \textsc{swift} \citep{Schaller2024} (smooth particle hydrodynamics (SPH) Lagrangian hydrodynamics with fast~multipole-particle~mesh gravity solver) code. For these simulations, initial conditions were generated for both dark matter and baryons (neglecting any baryon isocurvature perturbations so that the two initially perfectly trace each other). We assumed non-radiative evolution for the gas, disabling any extra astrophysics such as a UV background and feedback processes. For \textsc{gizmo} and \textsc{swift}, we set a comoving gravitational force softening of $2\,\text{kpc}$. For \textsc{ramses}, we enabled Lagrangian refinement with a maximum refinement level of 18, so that the smallest grid cells at $z=0$ have a linear size of $\sim0.4$~kpc.

In \autoref{fig:richings_bary4}, we show the resulting dark matter density, gas density, gas temperature, and gas entropy maps at the end of the simulations for each of the three codes. We find a generally good agreement of the large-scale structure visible in the three simulations, in both dark matter and gas density, as well as a good agreement in the thermodynamic properties of the gas inside the virial shock of the halo. While some differences exist in the thermodynamic properties of the gas, we leave a detailed discussion of them to code comparison studies, such as e.g. \citet{RocaFabrega2021}. Instead, we note that with \cosmicweb, it is easily possible to resimulate any object listed in the database with all of the state-of-the-art cosmological hydrodynamics, which we expect to greatly facilitate the reproducibility of research in this area.

\section{Conclusion}\label{sec:conclusions}

We have presented the \cosmicweb project, an online platform that aims to simplify the creation and sharing of zoom simulations. We have laid out the current work-flow for setting up and running zoom simulations and showed how \cosmicweb integrates with this process. In particular, we address three key steps: i) finding and selecting targets for zoom simulations, ii) retrieving the zoom initial conditions with a high resolution region placed on the Lagrangian volume of the target, and iii), sharing the initial condition configuration in research publication, allowing the community to compare and verify scientific results.
We envision the following workflow with \cosmicweb:
\begin{enumerate}
 \item select a suitable simulation for simulation target,
 \item choose appropriate filters in the \emph{Halo Finder} and query results,
 \item preselect some halos using the property list and scatter plot and store them in a new collection,
 \item take a closer look at individual halos in the \emph{Halo Explorer},
 \item select the target halo(s) by either removing halos from the collection or by creating a new collection,
 \item download the initial conditions configuration files for the halo(s) using the \emph{Downloader} overlay or the \music plugin,
 \item create the initial conditions with \music and run the zoom simulations,
 \item when publishing the results, promote the collection to a publication, include the generated \cosmicweb link in the article, and add a link to the article in \cosmicweb.
\end{enumerate}
The ability to access initial conditions from published research will allow the scientific community to easily verify results and compare and improve different models of baryonic physics and code implementations, which will ultimately help us to improve our overall understanding of the evolution of galaxies and clusters in our Universe.

The proto-halo patches offered by \cosmicweb are in the form of minimal bounding ellipsoid, which capture the shape of a typical proto-halo patch without requiring a large amount of storage. For each halo, we provide ellipsoids at varying traceback-radii. To evaluate the goodness-of-fit of these ellipsoids, thereby determining the computational overhead by high-resolution particles not part of the final target volume, we have provided an analysis of the ellipsoid efficiency parameters.  We find that the efficiency is significantly shaped by the halo's mass and environmental context. High-mass halos and those in less dense environments tend to have more spherical mass distributions, which are more accurately encapsulated by ellipsoids than those of lower mass halos in regions with strong tidal forces.

In the context of selecting targets for zoom simulations, choosing proto-halos that are well-described by their ellipsoids can reduce computational demands. Our Halo-Finder tool leverages ellipsoid efficiency as a criterion for selection. However, this may lead to a preference for isolated halos, potentially introducing a bias in the formation histories of the selected objects. This bias warrants careful consideration to ensure an unbiased representation in zoom-simulation studies.

We provide an updated version 2 of the \music\footnote{The updated version 2 of \music is available from \url{https://github.com/cosmo-sims/MUSIC}} initial condition generator software that supports \panphasia noise fields along with other improvements such as e.g., a reduced memory usage and direct interfacing with the \textsc{Class} linear Einstein-Boltzmann code.

Development of \cosmicweb is a continuing effort, since the data as well as the requirements change over time, requiring adaptations and extensions to the project. 
The kind of improvements that we will consider highly depends on the feedback and requirements from the community. 
We outline our plans in \autoref{sec:future_directions}.

\section*{Data availability}
The data underlying this article will be shared on reasonable request
to the corresponding author.

\section*{Acknowledgements}
MB and OH acknowledge funding from the European Research Council (ERC) under the European Union's Horizon 2020 research and innovation programme (grant agreement No. 679145, project 'COSMO-SIMS'). JCH and ARJ acknowledge support from UKRI grants ST/T000244/1 and ST/X001075/1.

This work used the DiRAC@Durham facility managed by the Institute for Computational Cosmology on behalf of the
STFC DiRAC HPC Facility (\url{www.dirac.ac.uk}). The equipment was funded by BEIS capital funding via STFC capital grants ST/K00042X/1, ST/P002293/1 and ST/R002371/1, Durham University and STFC operations grant ST/R000832/1. DiRAC is part of the UK National e-Infrastructure.
This work was granted access to the HPC resources of TGCC/CINES under the allocation A0060410847 attributed by GENCI (Grand Equipement National de Calcul Intensif).
Some of the computational results presented have been achieved using the Vienna Scientific Cluster (VSC).
\newpage
\bibliography{references}{}
\bibliographystyle{aasjournal}

\appendix 

\section{Future Directions}\label{sec:future_directions}
We categorize future developments into two classes: the addition of new data, in particular the integration of new simulations, and the addition of new features. 

\subsection{New simulations and data}
New halo catalogs and proto-halos from additional simulations can either be directly hosted on \cosmicweb servers, or from existing or new databases located outside. To facilitate the setting up of the API interface, we may publish our own implementation described in \autoref{sec:local_api} as a template in the future. 

Running zoom simulations on targets in the \cosmicweb database will naturally improve the knowledge on these objects, for example improving estimates of properties of the halo and the embedded galaxy and better constraining Lagrangian volumes of the proto-halo. 
In the future, we hope to provide a mechanism to feed this information back into the database and therefore provide more fine-grained information for later runs on these targets.

\subsection{Additional functionality}
As a result of the modular design, new exploration features and filters can be easily added in the future if required. Such features may include visualizations of the full halo merger tree with the ability to navigate different progenitor branches, animations of the halo growth, motion, and merger events using the already existing 3D framework for the visualization of the halo surrounding, or a tool to find similar halos in the database for statistical studies (e.g., by suggesting halos with similar properties).

For initial conditions generation and proto-halos, we plan to add support for alternative generators, such as the native implementation of \icgen \citep{Jenkins2010,Jenkins2013} in addition to the \music plugin. We may also consider supporting additional descriptors of proto-halo shapes, such as convex hulls, for more optimized zoom simulations.

\section{API Server Requirements}\label{sec:api_documentation}
The communication between the \cosmicweb application and external databases is handled by API Servers following a predefined set of rules. Since these specifications are subject to change over time, the outline given here reflects the state at the date of release. Up-to-date and more detailed specifications can be found directly on the webpage\footnote{\url{\cosmicweburl/documentation/api}}.

The API consists of several endpoints that receive requests from the client application and send the requested data back in the \texttt{JSON} format. We list and explain these endpoints in \autoref{tab:api_endpoints}.
Different simulations may not contain all data and support all features. This is why we differentiate between \emph{required} and \emph{optional} endpoints. The information on which features and properties are available for each simulation are stored on the central server.

\begin{table}
  \begin{threeparttable}
    \caption{List of API endpoints, split into required and optional features. The endpoints use the HTTP \texttt{GET} method with the exception of \texttt{/halo\_finder}, for which the selectors will be transmitted by a \texttt{POST} request. The endpoint paths are relative to the simulation API URL stored on the \cosmicweb application server. The precise format of the request and JSON response is specified on \href{https://cosmicweb.astro.univie.ac.at/documentation/api}{\cosmicweb}.\label{tab:api_endpoints}}
    \begin{tabularx}{\linewidth}{@{}l c X@{}}
      \toprule
      endpoint & request & \\
      \midrule
      \multicolumn{3}{@{}l}{\textbf{Required Endpoints}} \\
      \cmidrule(r){1-1}
      \texttt{/halo\_finder} & \post & Interface for halo queries, accepting filters to specify halo parameter ranges, mass definitions, availability of ellipsoids, etc. \\
      \texttt{/snapshots} & \get & List of available snapshots for the simulation, including redshifts and number of halos \\
      \texttt{/snapshots/<id>} & \get & Details of a specific snapshot such as redshift and number of halos \\
      \texttt{/halo/<id>} & \get & Details of a specific halo (and associated galaxy if available). The returned properties depend on the enabled features for the simulation (cf. \autoref{tab:features}). \\
      \texttt{/halo/<id>/ellipsoids} & \get & All Lagrangian ellipsoids for the halo (measured at different traceback-radii), specified by the center, shape matrix, and traceback-radius (optional: number of particles within $R_\mathrm{tb}$ and efficiency parameter). \\
      \texttt{/halo/<id>/ellipsoids/<id>} & \get & Definition of a specific Lagrangian ellipsoid \\
      \midrule
      \multicolumn{3}{@{}l}{\textbf{Optional Endpoints}} \\
      \cmidrule(r){1-1}
      \texttt{/snapshots/<id>/massfunction} & \get & The halo mass function computed from all host halos in the snapshot. \\
      \texttt{/halo/<id>/substructure} & \get & Substructure tree of the halo \\
      \texttt{/halo/<id>/main\_progenitors} & \get & List of halos corresponding to the main progenitors sequence of the specified halo \\
      \texttt{/halo/<id>/mergertree} & \get & Merger tree of the halo. The merger threshold can be specified by the request argument \texttt{?mm\_fraction=<r>}, where $r$ is the ratio between merging and resulting halo. A 1:3 merger for example corresponds to $r=0.25$. \\
      \texttt{/halo/<id>/surrounding} & \get & All halos within a certain distance of the halo. The distance can be specified by the request argument \texttt{?box\_size=<m>}, where $m$ is a multiplier to the halo radius. \\
      \texttt{/halo/<id>/tracking} & \get & A list of coordinates of the main progenitors, normalized to the unit cube \\
      \texttt{/halo/<id>/images} & \get & If there are images or graphics associated with the halo, return a list of image URLs and descriptions \\
      \bottomrule
    \end{tabularx}
  \end{threeparttable}
\end{table}

The required endpoints are necessary for the basic functionality of \cosmicweb. These include information on the available snapshots, the halo-finding capability, basic information on individual halos, and their Lagrangian ellipsoids.
The optional endpoints are mainly used for visualizations of the halo substructure, the formation history, and the local environment of the halo. For each simulation in \cosmicweb, individual features can be enabled, and the interface automatically adapts to show the available information.

External datasets can be integrated in \cosmicweb by setting up a server listening on the various endpoints, parsing incoming requests, querying the dataset, and returning the result in the specified format. For the \cosmicweb simulations, we are using the \textsc{flask} framework for that purpose, but in general any web server that can parse and process the requests is suitable.

\section{\cosmicwebmusic Command Line Interface}\label{sec:music_plugin}
To facilitate downloading initial conditions from \cosmicweb to computing facilities and local machines, we provide a \textsc{Python} command-line executable\footnote{\url{https://github.com/cosmo-sims/cosmICweb-music}} that interfaces with the \cosmicweb API. 
It allows downloading individual initial condition configuration files and bulk configurations for collections and publication lists. The executable has two modes (\texttt{publication} and \texttt{get}) and is used as:

\vspace{0.2cm}
\begin{description}
    \item[\texttt{cosmicweb-music}] \texttt{[--output-path OUTPUT\_PATH] [--common-directory]} \\
    \texttt{[publication  [--traceback-radius RADIUS] PUBLICATION\_TAG]} \\
    (or) \\ 
    \texttt{[get DOWNLOAD\_STRING]}
    \vspace{0.2cm}
    \item[\texttt{--output-path OUTPUT\_PATH}:] By default, the \music configuration files are stored in the current working directory in a sub-directory named after the publication or the simulation of the halos. If \texttt{--output-path} is specified, the script stores the files in this directory instead without creating the subdirectory. Folders for each halo will be created however, unless the \texttt{--common-directory} flag is set.
    \item[\texttt{--common-directory}:] If this option is set, all \music configuration files will be stored in the same directory instead of individual directories for each halo. 
    \item[\texttt{publication [--traceback-radius TRACEBACK\_RADIUS] PUBLICATION\_TAG}:] The tag of the publication can be used to download zooms for all halos within the publication list. The traceback-radius for the proto-halo patches is set to 2 as default but can be changed. The executable will skip a halo if no minimum bounding ellipsoid with the specified traceback radius exists.
    \item[\texttt{get DOWNLOAD\_STRING}:] A download string can be generated from the download interface on \cosmicweb. This string can then be used to download the \music configuration files with this executable. Configuration settings from the download window will be applied during the process. 
\end{description}

\begin{figure*}
    \includegraphics[width=0.97\textwidth]{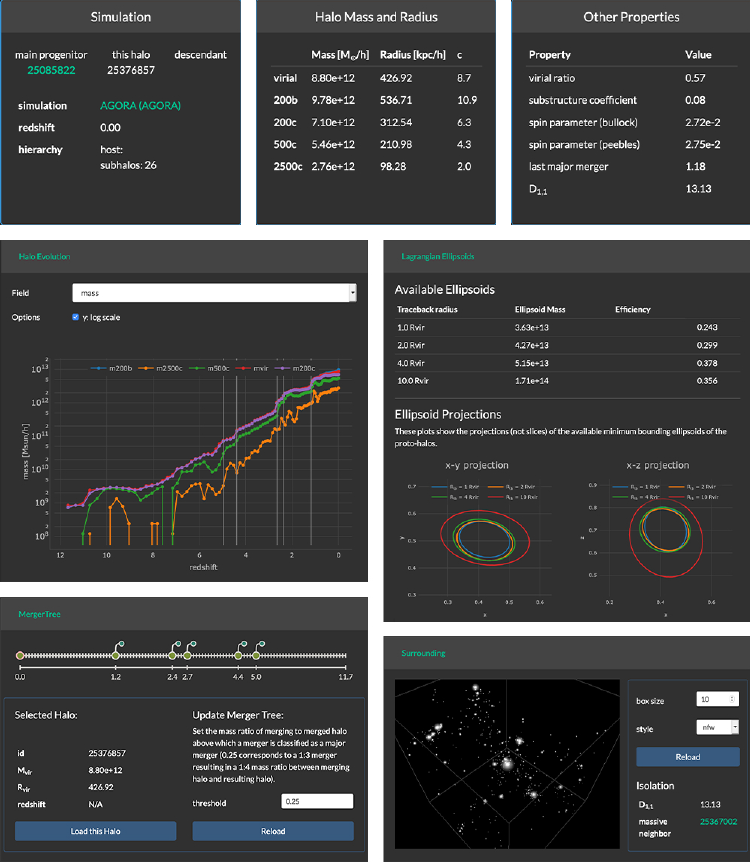}
    \caption{Screenshots of some visualization tools in the \emph{Halo Explorer}: overview of the halo parameters and links for main progenitor and descendant halo (top row), evolution of the halo parameters along the main progenitor branch (middle left), visualization of the major merger events with variable threshold (bottom left), list and projections of the Lagrangian minimum bounding ellipsoids (middle right), and a 3D WebGL rendering of halos in the surrounding.}
    \label{fig:screenshots}
\end{figure*}

\end{document}